\documentclass[fleqn,usenatbib]{mnras}

\usepackage{newtxtext,newtxmath}

\usepackage[T1]{fontenc}

\DeclareRobustCommand{\VAN}[3]{#2}
\let\VANthebibliography\thebibliography
\def\thebibliography{\DeclareRobustCommand{\VAN}[3]{##3}\VANthebibliography}

\usepackage{graphicx}	
\usepackage{amsmath}	
\usepackage{amssymb}	
\usepackage{geometry}
\usepackage{pdflscape}
\usepackage{xcolor}

\newcommand{\GAIA}{{\it Gaia}}
\newcommand{\xmm}{{\it XMM-Newton}}
\newcommand{\swift}{{\it Swift}}
\newcommand{\maxi}{{\it MAXI}}
\newcommand{\nustar}{{\it NuSTAR}}
\newcommand{\nicer}{{\it NICER}}
\newcommand{\mx}{MAXI J1820+070}
\newcommand{\Msun}{\hbox{$\rm\thinspace M_{\odot}$}}

\title[Investigating the Hard State of \mx]{Investigating the Hard State of \mx: A Comprehensive Bayesian Approach to Black Hole Spin and Accretion Properties}

\author[S. D. Dias et al.]{
Sachin D. Dias,$^{1}$\thanks{E-mail: sdd9@leicester.ac.uk}
Simon Vaughan,$^{1}$
Mehdy Lefkir, $^{1}$
Graham Wynn $^{1, 2}$
\\
$^{1}$School of Physics and Astronomy, University of Leicester, Leicester, LE1 7RH, UK \\
$^{2}$Department of Mathematics, Physics and Electrical Engineering, Northumbria University, Newcastle-upon-Tyne, NE1 8ST
}

\date{Accepted 2024 February 13. Received 2024 February 07; in original form 2023 November 28}

\pubyear{2023}

\begin{document}
\label{firstpage}
\pagerange{\pageref{firstpage}--\pageref{lastpage}}
\maketitle

\begin{abstract}
We analyse the X-ray spectrum of the black hole X-ray binary \mx\ using observations from \xmm\ and \nustar\ during 'hard' states of its 2018-2019 outburst. 
We take a fully Bayesian approach, and this is one of the first papers to present a fully Bayesian workflow for the analysis of an X-ray binary X-ray spectrum. This allows us to leverage the relatively well-understood distance and binary system properties (like inclination and black hole mass), as well as information from the \xmm\ RGS data to assess the foreground X-ray absorption.
We employ a spectral model for a `vanilla' disc-corona system: the disc is flat and in the plane perpendicular to the axis of the jet and the black hole spin, the disc extends inwards to the innermost stable circular orbit around the black hole, and the (non-thermal) hard X-ray photons are up-scattered soft X-ray photons originating from the disc thermal emission.
Together, these provide tight constraints on the spectral model and, in combination with the strong prior information about the system, mean we can then constrain other parameters that are poorly understood such as the disc colour correction factor. 
By marginalising over all the parameters, we calculate a posterior density for the black hole spin parameter, $a$. 
Our modelling suggests a preference for low or negative spin values, although this could plausibly be reproduced by higher spins and a modest degree of disc truncation. This approach demonstrates the efficacy and some of the complexities of Bayesian methods for X-ray spectral analysis.
\end{abstract}




\section{Introduction}
\label{Introduction}

Low mass X-ray binaries (LMXBs) comprise a compact object - either neutron star or black hole (BH) - accreting gas from a companion star. These powerful and exotic objects offer a view of the behaviour of matter under extreme conditions including in the strong gravity of a compact object. The compact object that is accreting is often called the \emph{primary}, and the lower-mass star providing the matter is called the donor, or \emph{secondary}. The gas flowing towards the primary will often form an accretion disc \citep{Shakura_Sunyaev_1973}, due to the conservation of angular momentum and viscous forces within the material that allow for the transfer of angular momentum outwards and mass inwards. Under certain conditions, the energy released by the gas flowing through the disc as it approaches the compact object can be a significant fraction of its rest mass energy. The disc becomes  hot and luminous -- with temperatures of $T \sim \text{few}\ 10^6$ K and luminosities up to $L \sim 10^{39}$ erg s$^{-1}$, making LMXBs among the brightest X-ray sources in the Galaxy. See \citet{McClintock_Remillard_2006} or \citet{Done_2007} for a review. 

Many mysteries remain about the inner accretion flow, where most of the energy is released. One outstanding problem is how close the putative accretion disc extends towards the compact object. This will be determined by a combination of the properties of the compact primary (including the spacetime metric in which the inner accretion flow moves) and the physical conditions in the hot accreting gas in the inner accretion disc. 

LMXBs tend to show intermittent periods of high X-ray luminosity (called \emph{outbursts}) between extended periods of low luminosity (called \emph{quiescence}). During outburst, LMXBs show two main types of X-ray spectra (but with some intermediate cases and exceptions): a hard spectral state and a soft spectral state. 
In the soft state, the X-ray spectrum is dominated by the thermal emission of the accretion disc, with a weak non-thermal spectrum extending above $10$ keV as a power law. In contrast, the hard state shows a spectrum dominated by the non-thermal (power law) component, but also sometimes features associated with X-ray `reflection'. The physical origin of the hard X-ray, non-thermal spectrum remains unclear, although it is most commonly explained in terms of inverse-Compton scattering of seed photons (original emitted by the hot disc) by a population of higher energy electrons in a `corona'. See \citet{Thorne_Price_1975} or \citet{Svensson_1994}.

In the case of LMXBs with a black hole primary, the mass ($M$) and spin (dimensionless angular momentum parameter $a = J/M^2$) define the spacetime in which the accretion flow exists. Of particular importance is the radius of the innermost stable circular orbit (ISCO). For a nonspinning black hole $R_{ISCO} = 6$ $r_g$ (where $r_g = GM/c^2$ is the gravitational radius of the black hole). For a black hole with high spin ($a=0.998$ is thought to be the maximum that can be physically realised; \citealp{Thorne_1974}), and orbits prograde with the spin, the ISCO moves into to $R_{ISCO} \sim 1$ $r_g$, whereas for a retrograde orbits around a maximally spinning black hole, $R_{ISCO} \sim 9$ $r_g$. The spin of black holes in LMXBs may be important in setting a limit to the inner edge of the accretion disc and also play a fundamental role in the production of relativistic jets that are often seen from XRBs and also their supermassive counterparts, the active galactic nuclei \citep{Blandford_1977, Steiner_2013}. The spin of black holes may provide insight into their formation in supernovae and subsequent evolution through accretion \citep{OShaughnessy_2017, Farr_2018}. Although the spin of the black hole cannot be directly observed, if the location of the inner edge of the accretion disc ($R_{in}$) can be estimated and associated with $R_{ISCO}$, one can then infer (subject to some assumptions) the spin parameter, $a$. The key observational challenge here is to estimate $R_{in}$ of the disc. 

There are currently three approaches used to estimate $R_{in}$. See \citet{Reynolds_2021}. The first method uses the shape of the X-ray reflection features in the spectrum, especially the profile of strong emission lines such as Fe-K$\alpha$ \citep{Fabian_1989}. If produced by florescence or resonance emission in the surface layers of the accretion disc, the shape of the emission line is affected by the radial distribution of the gas, determined in part by $R_{in}$. A broader profile is produced by a disc than extends closer to the black hole. This method is sometimes known as the 'reflection method'. 

The second method involves modelling the spectrum of the thermal emission of the accretion disc, which is also affected by the radial distribution of the disc material and hence $R_{in}$. Roughly, a disc that extends closer to the black hole tends to be both more luminous and hotter, all other things being equal. This is known as `continuum fitting' \citep{Zhang_1997, McClintock_2011}. 

These spectral effects can be subtle and are often difficult to disentangle. For instance the strength of the Fe-K$\alpha$ line used in reflection modelling is a combination of the relative strength of the reflected emission (which is connected to the geometry of the source and reflector), the ionisation factor and the Fe abundance. These effects are difficult to separate and so the line strength alone cannot determine the strength of reflection \citep{Garcia_2011}. 
Both the reflection and thermal spectrum of the disc will be affected by the inclination of the disc, and by radiative transfer effects in the disc atmosphere that are poorly understood. The profile of emission lines in a reflection spectrum also depend on the way in which the hard X-rays illuminate the disc \citep{Dauser_2013}. For reasons such as these, it is not possible to obtain simple estimates of $R_{in}$ based on the observed shape of emission lines; detailed modelling work is required. 

A third method involves identifying the frequencies of certain quasi-periodic variations in the X-ray luminosity and associating them with particular orbital perturbations possible in the space-time around a black hole. This method relies on assumptions very different from those in the two methods based on X-ray spectra discussed above. See e.g. \citet{Motta_2014}.

There is currently no consensus on the geometry of the accretion disc through outburst. The difficulties or ambiguities of modelling these spectra mean that different groups can reach conflicting conclusions, often based on the same data (see Section \ref{black_hole_spin}). 
Broadly speaking, there are two main lines of thought about the placement of inner edge of the (standard, thin) accretion disc.
\begin{enumerate}
 \item The standard, optically thick disc extends close to the black hole (R$_{in}$ $\approx$ R$_{ISCO}$) throughout all states of an outburst \citep{Reynolds_2021}. These models are supported by observations of both the Fe-K$\alpha$ peak \citep{Miller_2015, Buisson_2019} and lags \citep{Kara_2019}. The radius of the inner edge of the disc in any state can then act as a proxy for $R_{ISCO}$ (and hence be used to infer $a$). The dramatic changes between the `states' may be due to changes in the energetics of the corona and/or jet. 
 
 \item In the harder states, particularly at lower luminosities, the disc 
 is truncated, meaning $R_{in} \gg R_{ISCO}$, \citep{Done_2007, Kolehmainen_2014,Tomsick_2009}. As above, in the softer, thermal-dominated state, a standard disc extends inwards, close to $R_{ISCO}$, But in harder states the inner region of the optically thick disc (between $R_{in}$ and $\sim R_{ISCO}$) is replaced by a geometrically thick, optically thin flow responsible for the hard X-ray emission. In these models, the hard state spectrum is expected to be less sensitive to the black hole spin because the standard disc ends outside of the region strongly affected by this parameter. 
 \end{enumerate}

We address these issues with a Bayesian approach to the analysis of XRB spectra. In this paper, we used a model without truncation to test whether such a model with this assumption can fit the data. We make explicit assumptions about the basic properties of the binary system and its accretion disc, which we call the {\it vanilla} model.
In particular, here we assume a flat, thin accretion disc with the axis of rotation aligned with the black hole spin axis (and the radio/X-ray jet inclination serves as a proxy for this) that locally emits a blackbody spectrum and extends inwards as far as $R_{ISCO}$ (i.e no truncation), with a zero torque inner boundary condition. The hard X-rays are produced by inverse-Compton scattering of seed photons from the disc by hot electrons (in a `corona'). These, in turn, reflect off the ionised surface of the disc. This framework provides some powerful constraints on the parameters of the model, conditional on our basic physical assumptions about the system. Additionally, we make use of strong prior information in our modelling of the broad-band X-ray spectrum. 
\cite{Eckersall_2017} demonstrated the importance of accurate modelling of X-ray absorption by the interstellar medium (ISM) for modelling XRB spectra. We use the \xmm\ RGS data to provide information about the X-ray absorption. Additionally, we use the best measurements of the binary system parameters. 

Combining all this information in a Bayesian framework provides advantages.
Among these are that parameters remain consistent with prior information about the binary system, and this can also reduce degeneracies between parameters. Most X-ray spectral modelling to date does not account for prior information on parameters, and its uncertainty, where this is available.
We focus on a model without truncation of the inner disc, or with very limited truncation, to investigate whether such a model is sensible and study its implications on the black hole spin.
As a test case for our approach we examine three observations of the LMXB \mx\ (also known as ASASSN-18ey) taken during its hard state. 

\begin{figure}
	\includegraphics[width=\columnwidth]{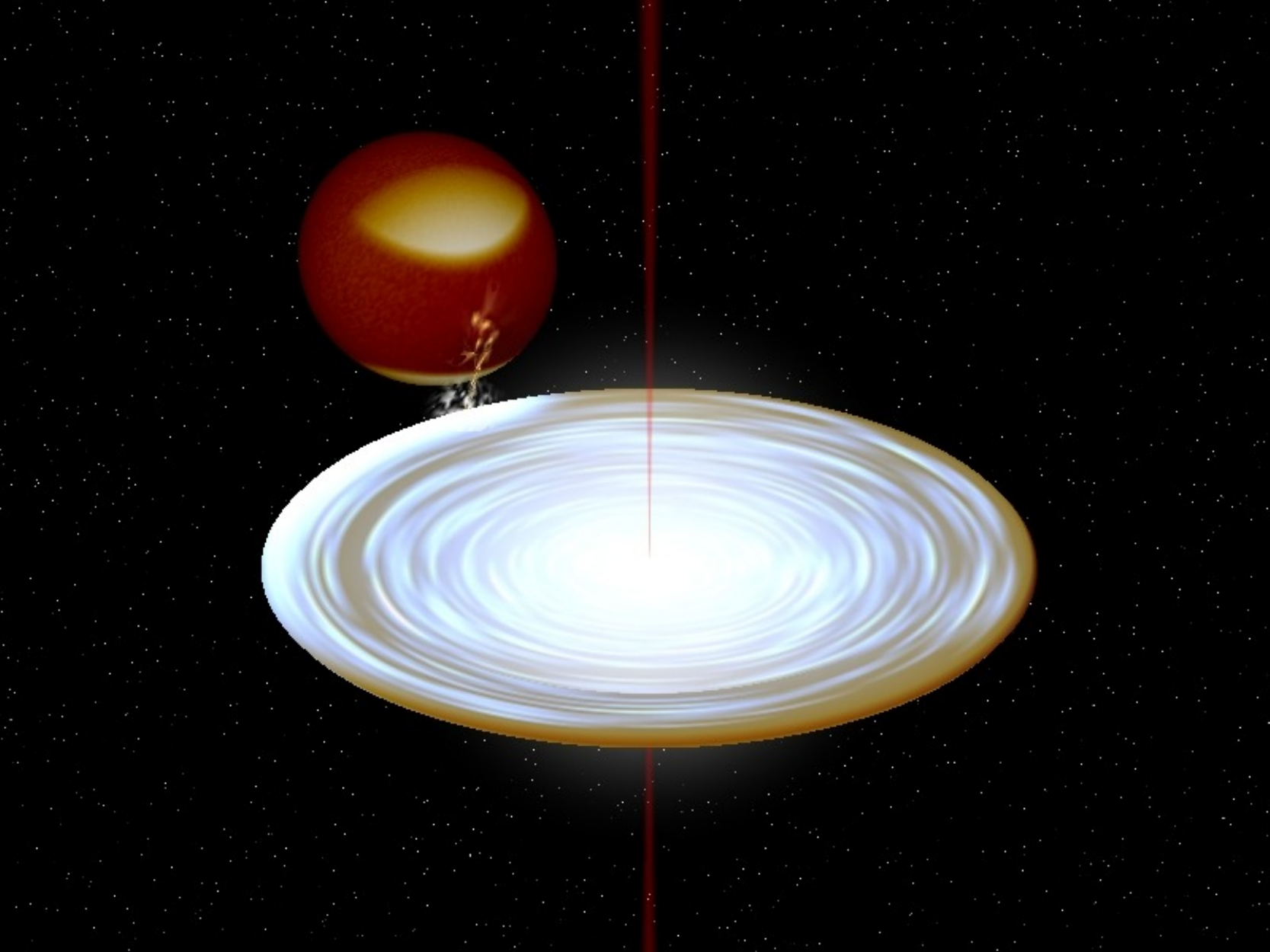}
    \caption{Simulated view of \mx\ made using \texttt{BINSIM} \citep{Hynes_binsim}. Input parameters are discussed in Section~\ref{MAXI_J1820}.}
    \label{fig:MAXI_BINSIM}
\end{figure}

The rest of this paper is organised as follows. 
The next section describes our data reduction for \xmm\ and \nustar\ observations and gives an overview of our approach to the spectral analysis.
In Section \ref{MAXI_J1820} we review some of the properties of the \mx\ system that are used to define prior densities on the binary system parameters.
Section \ref{RGS_section} describes modelling of the high-resolution soft X-ray data from the RGS. These data are used with a simple continuum model to constrain the ISM absorption column densities for neutral H, O, Ne and Fe as well as some ionised species. The results are used to define prior densities for the ISM absorption model. 
Then Section \ref{Vanilla_model} outlines our broad-band model, along with the definition of the remaining priors and their justification. 
Following this, Section \ref{EPIC_pn_NuSTAR} takes the model and applies it to the EPIC pn and \nustar\ data of \mx.
We then discuss the implications of our results in Section \ref{disucssion}.
Throughout this paper we make use of both natural and base-$10$ logarithms, indicated by the $\ln(\cdot)$ and $\log(\cdot)$ functions, respectively. 

\section{Observations and data reduction} \label{Obs&Dat}

\begin{table*}
	\centering
	\caption{Observation log. 
Column $1$ gives the name by which we refer to the three sets of observations, column $2$ gives the mission that data is taken from. Column $3$ gives the Observation ID number, column $4$ gives the revolution number in the case of \xmm\ observations. 
Column $5$ gives the observation start date in Julian date and MJD, column $6$ gives exposure times; these are given for both FPM-A and FPM-B in the case of \nustar\ data, and for EPIC pn in the case of \xmm\ data. Column $7$ gives count rates: for \nustar\ data, both FPM-A and FPM-B rates are given, and for \xmm\ data count rates are stated for both RGS and EPIC pn.
Column $8$ states whether or not the central column of the EPIC pn image was removed to mitigate small levels of pile up.}
	\label{tab:OBS_TABLE}
	\begin{tabular}{llccccccc}
		\hline 
		\hline
		Obs & Mission & Obs. ID & Rev. & Start Date (MJD) & Exposure (ks) & Count Rate (ct s$^{-1}$) & Pile Up?\\
		(1) & (2) & (3) & (4) & (5) & (6) & (7) & (8)\\
		\hline
		Obs1 & \xmm    & 0844230201  & 3531 & 2019-03-22 (58564) & 8.49        & 5.09/151.3  & Y\\
        Obs2 & \nustar & 90501311002 & -    & 2019-03-25 (58567) & 28.66/28.59 & 16.53/15.12 & -\\
		Obs2 & \xmm    & 0844230301  & 3533 & 2019-03-26 (58568) & 11.34       & 4.28/127.7  & Y\\
		Obs3 & \nustar & 90501337004 & -    & 2019-09-20 (58746) & 47.62/47.32 & 1.35/1.27   & -\\
		Obs3 & \xmm    & 0851181301  & 3623 & 2019-09-20 (58746) & 56.28       & 0.47/15.06  & N\\
		\hline
		\hline
	\end{tabular}
\end{table*}

\subsection{\xmm} \label{xmm_newton}

\begin{figure*}
	\includegraphics[width=16cm]{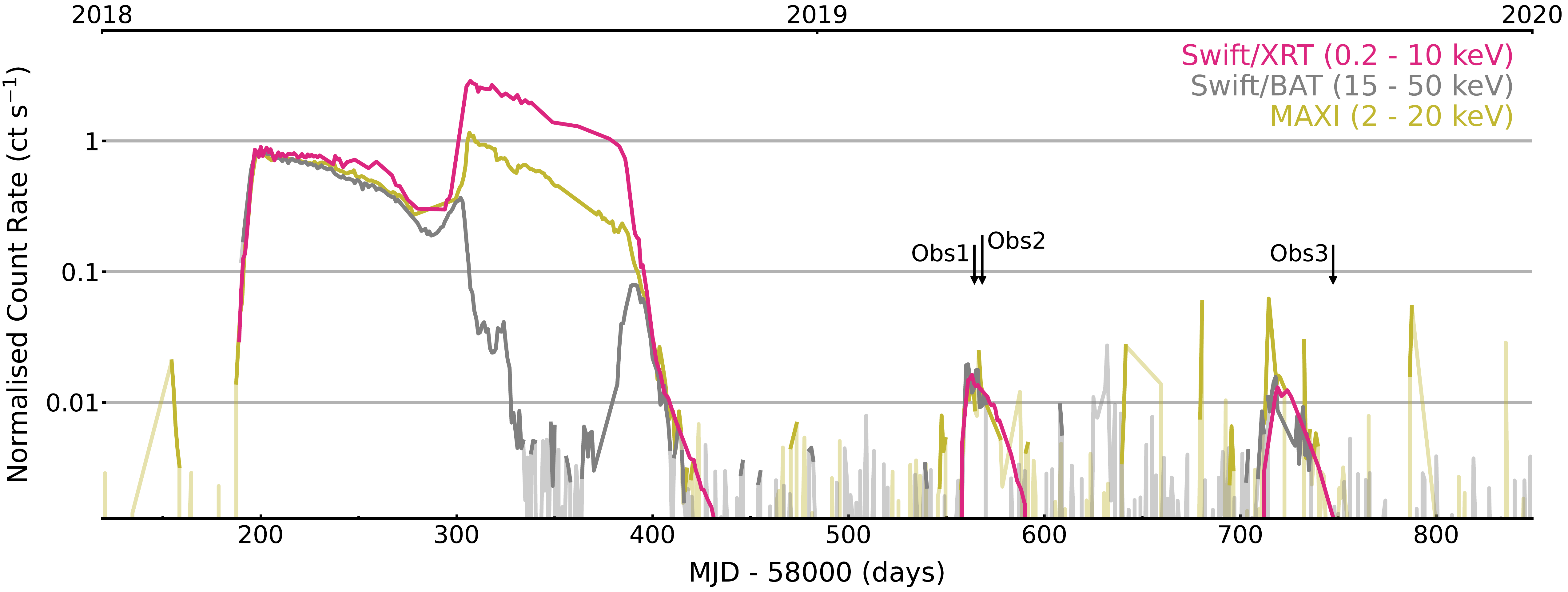}
    \caption{The \swift/XRT (pink), \swift/BAT (grey) and \maxi\ (yellow) light curves of MAXI J1820+070. Faded and opaque lines represent signal-to-noise below and above one respectively. The positions of \xmm\ observations are represented by arrows. }
    \label{fig:DATES_vs_FLUX}
\end{figure*}

We used three \xmm\ observations of \mx\ taken in March and September 2019, all while the system was in a hard state.
These observations occurred months after the initial outburst had faded, during two separate 're-brightening' episodes (mini-outburst/reflares; \citealp{Xu_2020, Ozbey_Arabac_2022}), as shown in Fig. \ref{fig:DATES_vs_FLUX}. In the rest of this paper we refer to these as obs1, obs2 and obs3 (see Table \ref{tab:OBS_TABLE} and Fig. \ref{fig:DATES_vs_FLUX}). 
There are several previous \xmm\ observations. We do not make use of these as in all cases the count rates were much higher and the EPIC data suffer significantly from pile up.

The observations were obtained from the \xmm\ Science Archive\footnote{\url{http://nxsa.esac.esa.int/nxsa-web/\#home}} and processed with the \xmm\ Science Analysis System (\texttt{SAS v16.0.0} for RGS data; \texttt{SAS v20.0.0} for EPIC pn data). 

The RGS were operated in \texttt{SPECTROSCOPY HER} (high event rate) mode + \texttt{SES} (Single Event Reconstruction) -- this limits the telemetry if the count rate was very high. The RGS spectrum and background files were obtained using {\tt RGSPROC v1.34.7}, with response matrices generated using {\tt RGSRMFGEN v1.15.6}. For each observation we produced a single, merged spectrum (source and background) combining data from RGS1 and RGS2 using {\tt RGSCOMBINE v1.3.7}. The RGS spectra were not binned, and for spectral fitting we used the $W$-stat likelihood function, to provide the maximum resolution of absorption features. The range $11 - 24$ \AA\ was chosen for these spectra as it included important absorption features from Ne, O and Fe. We ignored the region $22.7 - 23.2$ \AA\ when fitting the spectra in order to exclude the strongest features caused by dust. As outlined in \cite{Pinto_2010} and \cite{Pinto_2013} there can be a significant opacity in this range from non-gaseous ISM components such as ice or silicates.

The EPIC pn event files were processed with the meta-task {\tt EPPROC v2.25.1}. {\tt EVSELECT v3.71.1} was then used to extract an image file in {\tt RAW} coordinates and examined with {\tt DS9 v7.6}. All three observations were taken in {\tt TIMING} mode where the EPIC pn detector captures a one-dimensional image of the source with a time resolution of $0.03$ ms \citep{EPIC_pn_Struder_2001}. The maximum count rate for this mode is $800$ ct s$^{-1}$\footnote{\url{http://xmm-tools.cosmos.esa.int/external/xmm\_user\_support/documentation/uhb/epicmode.html}}, and so we expect a small level of event `pile up' in these exposures. For obs1 and obs2, some exposure time was in {\tt BURST} mode -- a variation on the {\tt TIMING} mode that can handle a much higher maximum count rate with minimal pile up. We chose not to do this as {\tt BURST} mode has a live time of only $3$\%, so the signal/noise will be much lower than for the {\tt TIMING} mode data. For each observation, a source region was selected with a width of approximately $19$ pixels centred around the brightest column. Corresponding background regions were chosen far from the source region, to minimise contamination from source photons, with a width of $15$ pixels. Both observations from March experienced small degrees of pile up and as such the central, brightest column was removed. The EPIC MOS spectra were ignored as the high count rate will cause severe pile up.

In order to extract the source and background spectra, {\tt EVSELECT} was used to extract events from the regions outlined above, selecting only events with {\tt FLAG == 0} and {\tt PATTERN <=4}. Redistribution matrix files and ancillary files were generated with {\tt RMFGEN v2.8.5} and {\tt ARFGEN v1.102.1}, respectively. Lastly, {\tt SPECGROUP v1.7.1} was used to re-bin the source spectrum such that each bin contains a minimum of $25$ counts and is not narrower than $1/3$ of the full width half maximum (FWHM) energy resolution of the central channel of the bin \citep{Kaastra_2016}. (As discussed later, the EPIC spectral modelling is computationally demanding, and some amount of spectral binning was necessary to speed up the process.) 

The energy range of the EPIC pn spectra was restricted to $0.8 - 10$ keV. Above $10$ keV the signal-to-noise ratio of the data decreases, and we found no improvement to the quality of our fits extending the data to $0.7 - 10$ keV. As is common with EPIC spectral analysis, we excluded the $1.75 - 2.35$ keV range to avoid residuals from the instrumental Si-K and Au-M edges \citep{EPIC_pn_Struder_2001, EPIC_MOS_Turner_2001}.

\subsection{\nustar}

We used two \nustar\ observations, simultaneous with \xmm\ observations from obs2 and obs3, see Table \ref{tab:OBS_TABLE}. The standard observing scientific mode $01$ was used for all observations. The data were obtained from the \nustar\ Archive \footnote{\url{https://heasarc.gsfc.nasa.gov/docs/nustar/nustar_archive.html}} and processed with {\tt NuSTARDAS v2.1.1} and with the use of \nustar\ {\tt CALDB v20220301}. Event files were created with the {\tt nupipeline}, with spectra then generated by {\tt nuproducts}. For each of the Focal Plane Module Detectors (FPM-A and FPM-B) $180$" source and background regions were identified using {\tt DS9 v4.1}, with the background placed so as to occupy a region of low source counts. In order to re-bin the spectra the {\tt FTOOLS v6.30} \citep{FTOOLS_2014} command {\tt ftgrouppha} was used with group type of {\tt optmin} and a {\tt groupscale} of $25$. This ensured spectra were binned following the optimal binning expression from \cite{Kaastra_2016}, with no fewer than $25$ counts per bin, similar to the approach used with the EPIC data (above). We fitted \nustar\ spectra over the energy range $3 - 70$ keV. Above $\sim$ 70 keV the source spectrum had dropped significantly below the background. In the case of the simultaneous observation with obs2, we further restricted the energy range of FPM-A to $5 - 70$ keV. This was due to a tear in the insulation of FPM-A, resulting in an over-abundance of events below $8$ keV, seen as a significant deviation from the spectrum of FPM-B \citep{Madsen_2020}. For this observation however, we found that above $5$ keV there was good agreement between FPM-A and FPM-B.

\subsection{Analysis Methods}

We use Bayesian methods -- where prior information on model parameters is combined with information from the data (through a likelihood function) to form a posterior density over all parameters. 

For the parameter inference problem, Bayes' rule says:
\begin{equation}
    p(\theta | D, M) = \frac{p(D|\theta, M)p(\theta|M)}{p(D|M)} = 
    \frac{\text{likelihood} \times \text{prior}}{Z}
\end{equation}
where $\theta$ are the parameters of a model $M$, and $D$ represents the data. The term $p(D|\theta, M)$ is the \emph{likelihood} for data $D$ as a function of parameters, $\theta$, and the term $p(\theta|M)$ is the \emph{prior}, encoding information about the parameters known prior to inspection of the data $D$. The denominator term, $p(D|M)$ is called the \emph{marginal likelihood} or \emph{evidence} and is sometimes denoted by $Z$. On the left side of the equation is what we wish to calculate: the \emph{posterior} distribution for the parameters given the data and all prior information. 
If we are only interested in parameter inferences, we can ignore the evidence term, which is constant with respect to $\theta$. 
Note that these terms are conditional on the choice of model, $M$, but they can be computed for other models and the parameters or evidence compared between choices of model. 

Our choice of priors is discussed over the next few sections. 
As is the case in most Bayesian analysis problems, the posterior cannot be computed analytically, and so sampling from the posterior is required to reveal its shape. Although we do optimise to find a `point estimate' (best fit) for the parameters, we consider the full shape of the posterior to draw our inferences. 

All spectra were analysed using  {\tt XSPEC v12.12.1a} \citep{Arnaud_1996}. For posterior sampling, we employed two different methods: Markov chain Monte Carlo (MCMC; see Appendix \ref{MCMC_Appendix}) and nested sampling (NS; see Appendix \ref{NS_Appendix}). 
For MCMC, we used the Goodman-Weare sampler \citep{GW_2010}. 
For NS, we used {\tt BXA v4.0.2} \citep{Buchner_2014} and {\tt UltraNest}\footnote{\url{https://johannesbuchner.github.io/UltraNest/}}{\tt v3.2.0} \citep{Buchner_2021}, with {\tt PyXspec}\footnote{\url{https://heasarc.gsfc.nasa.gov/xanadu/xspec/python/html/index.html}} {\tt v2.1.0} and {\tt Python v3.7.4}. 

For the RGS analysis, we require maximum spectral resolution to retain information about narrow absorption features. We did not rebin the data (retaining the default resolution) and used a Poisson likelihood function suitable for low counts per bin data ({\tt cstat} in {\tt XSPEC}). 

For the combined EPIC and \nustar\ analysis we used rebinned data in order to improve the speed of the samplers relative to unbinned data. With moderate-high counts per bin, the Poisson likelihood should become approximately Gaussian, and so using a Gaussian likelihood function ({\tt chi} in {\tt XSPEC}) should give approximately the same results as using a Poisson likelihood function ({\tt cstat} in {\tt XSPEC}). As a first check, we simulated spectra using a model based on preliminary fits to the EPIC data and fitted these using each statistic, finding that sampling using a $\chi^2$ likelihood was reasonably efficient and returned parameters with very little bias. We used {\tt chi} with the MCMC sampler, and {\tt cstat} with the NS sampler, deliberately using two different likelihoods and two different samplers as a check for consistency of the methods. 

For the model comparison, we focus on the Bayes factor (BF) computed in two different ways. From the MCMC and NS output we compute the Savage-Dickey Density Ratio (SDDR) which gives the BF for nested models. Nested sampling returns a direct estimate of the marginal likelihood (aka {\it evidence}) for each model, the ratio of marginal likelihoods for two models gives the BF. See Appendix \ref{Analysis_Methods}.

\section{The properties of \mx} \label{MAXI_J1820}

\mx\ was discovered during an outburst in March 2018 \citep{Denisenko_2018} by the All-Sky Automated Survey for SuperNovae (ASAS-SN; \citealp{Shappee_2014}). A few days after the initial optical detection, it was detected in X-rays \citep{Kawamuro_2018, Kennea_2018} by the Monitor of All-sky X-ray Image (\maxi; \citealp{Matsuoka_2009}), and the \swift\ Burst Alert Telescope (BAT; \citealp{Krimm_2013}). To date, the system has shown only this one outburst, continuing for approximately a year after discovery, followed by lower level activity. Following its discovery, \mx\ displayed a rapid rise to a flux of $\sim 2$ Crab in X-rays ($2 - 20$ keV; \citealp{Shidatsu_2018, Xu_2020}) while remaining in a hard spectral state until July 2018 when it transitioned into a soft state dominated by a thermal X-ray spectrum \citep{Homan_2018, Shidatsu_2019} with reduced radio through infrared emission \citep{Shidatsu_2019, Casella_2018, Tetarenko_2018}. It remained in a soft state for approximately three months, before transitioning through an intermediate state to the hard state \citep{Shidatsu_2019} and on into quiescence, where it remains to this day. During outburst, the source was bright and in a direction of relatively low Galactic absorption, making it an excellent target for multiwavelength observations from radio through to gamma-rays.

\subsection{Binary system measurements}
\label{sect:binary}

\cite{Espinasse_2020} were able to observe the emission of relativistic X-ray jets, confirming the source as a microquasar. 
The mass function (giving a lower limit on the mass of the primary) was estimated to be $\approx 5$ \Msun\ using dynamical methods by  \cite{Torres_2019}, establishing its black hole nature. Combining this mass function with a mass ratio ($q = M_2/M_1$) $\approx 0.12$ and an inclination of $69^{\circ} \lesssim i \lesssim 77^{\circ}$ (from the detection of a grazing eclipse), they estimated a black hole mass in the range $7-8$ \Msun. 

\citet{Bright_2020} detected the proper motion of jet ejecta, both approaching and receding, using radio imaging. Further radio measurements with the Very Long Baseline Array (VLBA) and the European Very Long Baseline Interferometry Network (VLBI) gave a parallax measurement of $0.348 \pm 0.033$ mas, giving a distance of $D = 2.96 \pm 0.33$ kpc \citep{Atri_2020}. Combining this with the proper motion of the ejecta allowed for a more precise inclination estimate of $i = 63 \pm 3^{\circ}$ \citep{Atri_2020}. In turn, these allowed for improved mass ratio ($q = 0.072 \pm 0.012$) and primary mass estimates ($M_1  = 8.48^{+0.79}_{-0.72}$ \Msun) \citep{Torres_2020}.

\subsection{System parameter priors}

We used these measurements of $M_1$, $q$, $i$ and $D$ to define prior densities for key system parameters of \mx\ to be used in our spectral modelling. 

\subsubsection{Distance, $D$}

For the distance prior, $p(D)$, we use the method outlined in \cite{BailerJones_2015} (Eqn. 1). This defines a measurement model for the parallax that can be used as a likelihood function for the distance. Combined with a prior for the distance, this gives a posterior density for the distance that we can then use as the prior for our X-ray spectral modelling. For the parallax measurement, we use the radio estimate from \cite{Atri_2020}, which is more precise than that from the \GAIA\ (Data Release 2) data given in \cite{Gandhi_2019}.
The resulting likelihood function is quite narrow, meaning that the posterior shape is relatively insensitive to the choice of the space density prior. Here, we use the exponential model described by \cite{Gandhi_2019}, with a scale length of $2.17 \pm 0.12$ kpc. In order to provide a simple description of the resulting distribution, we evaluate the posterior on a fine grid of distances, and then model this using a $\Gamma$ density function. In this case, the result is well-approximated by a $\Gamma$ density with a peak probability density at $\sim$ 3 kpc, shape parameter $\alpha = 31.66$ and inverse scale parameter $\beta = 10.61$ (see Fig.~\ref{fig:farr_method}).

\subsubsection{Inclination, $i$} \label{sys_params_inclination}

Our prior for the accretion disc inclination was based directly on the inclination estimate of $i = 63 \pm 3^{\circ}$ from \cite{Atri_2020}. We assume the inner disc is perpendicular to the jet and so the estimates of $i$ based on the jet double as estimates of the inner disc inclination. We defined a prior $p(i)$ using a Gaussian with mean $\mu = 63$ and standard deviation $\sigma = 3$.

We note in passing that \citet{Poutanen_2022} argues for a 40$^{\circ}$ misalignment between jet and binary plane in \mx.

\subsubsection{Black hole mass, $M_{BH}$}

We constructed a prior for the black hole mass $p(M_1)$ using the method discussed in Section $2$ of \cite{Farr_2011}. 
In outline, we use distributions for $i$, $q$ and the mass function $f(M_1, M_2)$ to build a density for $M_1$. Given estimates for the period of the binary orbit ($P$) and amplitude of the periodic radial velocity variations of the secondary star ($K$), from e.g. optical spectroscopic monitoring, one can compute $f(M_1, M_2)$ as follows:
\begin{equation}
    f(M_1, M_2) = \frac{K^3 P (1-e^2)^{3/2}}{2 \pi G} = \frac{(M_1 \sin(i))^3}{(M_1 + M_2)^2}.
\label{eq:mass_function_1}
\end{equation} Here, $M_2$ is the mass of the secondary star, $e$ is the eccentricity of the orbit ($0 \le e < 1$), and $G$ is the gravitational constant. From the spectrum and and rotational velocity of the secondary star, the mass ratio ($q= M_1/M_2$) can be determined. The mass function (Eqn. \ref{eq:mass_function_1}) can be expressed in terms of $M_1$ and $q$: 
\begin{equation}
    M_1=\frac{f(M_1, q)(1+q)^2}{\sin^3(i)}.
\label{eq:mass_function_2}
\end{equation} 
We assign a probability distribution to each of $f$, $q$ and $i$, then use these to generate sets of random numbers, and for each set ($f$, $q$ and $i$) we compute a mass, $M_1$. The result is a distribution of $M_1$ estimates accounting for observational uncertainties.
For $f$, we use $5.18 \pm 0.15$ \Msun\ from \cite{Torres_2019} as this is the only dynamical estimate in the literature to date. 
We use the $q = 0.072 \pm 0.012$ from \cite{Torres_2020}, and for the inclination $i = 63 \pm 3^{\circ}$ (as discussed above). As each is reasonably well determined, we model the uncertainties using Gaussian densities on all three parameters. 

\begin{figure*}
	\includegraphics[width=16cm]{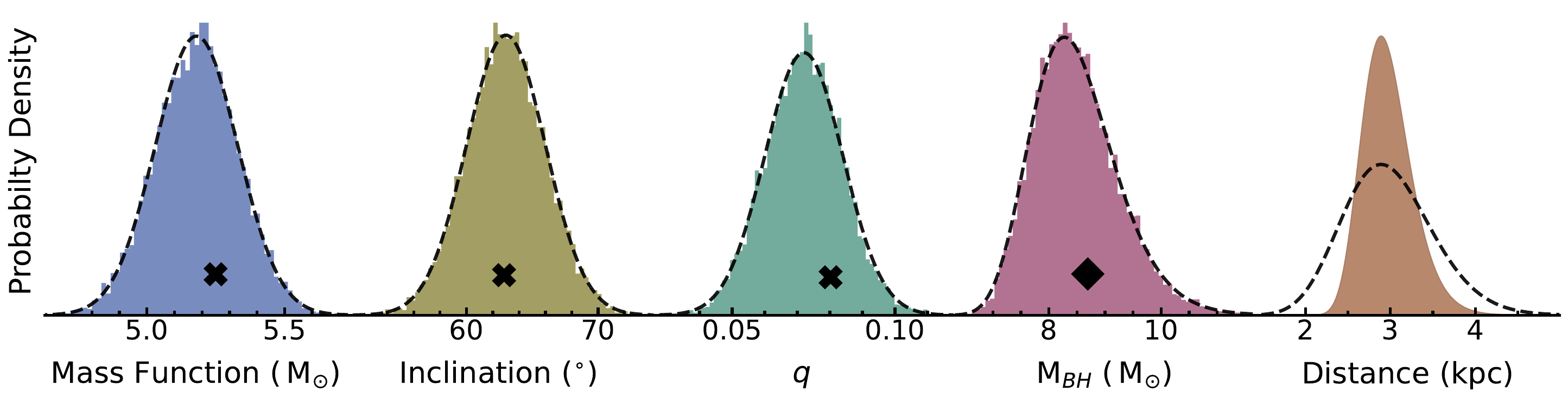}
    \caption{Posterior distributions for the mass function, inclination, mass ratio and black hole mass using the method outlined in \protect\cite{Farr_2011} and distance using the method outlined in \protect\cite{BailerJones_2015}. In the first three histograms, a black dotted line represents the priors used to generate these distributions. For black hole mass, the black dotted line represents the shifted log-normal probability density function which we approximate this distribution to be. This will become our prior on the mass of the black hole (see Section \ref{EPIC_pn_NuSTAR}). Black crosses show one random draw of the mass function, inclination and mass ratio, with the corresponding $M_{BH}$ these values derive shown by a black diamond. Similarly, for distance, the black dotted represents the gamma probability density function which we approximate this distribution to be and use as our distance prior (see Section \ref{EPIC_pn_NuSTAR}). }
    \label{fig:farr_method}
\end{figure*}

From these distributions, we sampled $N = 2 \times 10^4$ estimates of $M_1$ (see Fig.~\ref{fig:farr_method}), and excluded all samples outside of the range $[2, 50]$ \Msun. We fit the resulting histogram with model based on a shifted log-normal density function, which we used as the prior $p(M)$. It has a shift of $4.49$ \Msun, $\mu = \ln(3.92)$ and $\sigma = 0.19$. 

The results of the above analysis are models for the prior distribution of key system parameters $D$, $i$ and $M_1$. For the spectral modelling (discussed later) we treat these as independent, i.e. the joint prior $p(D, i, M_1)$ is the simple product of the individual (marginal) priors. As the mass estimate depends on the inclination, these two are not independent but treating them as independent is in a sense more conservative as the joint density is spread over a larger volume of parameter space.

\subsection{Black hole spin} 
\label{black_hole_spin}

As with many XRBs, there is a considerable debate over both the spin of the black hole and the radius of the inner edge of the accretion disc in MAXI J1820+070. 

Several previous studies have attempted to constrain or estimate the black hole spin in \mx.

Using the continuum fitting method with soft state data, \citet{Guan_2021} and \citet{Zhao_2021} found low spin, $a \lesssim 0.2$.
By contrast, \citet{Bharali_2019} used X-ray reflection modelling of hard state data from \nustar\ and \swift\ and found $a>0.68$. \citet{Draghis_2023} also used reflection modelling of \nustar\ data and found $a=0.98^{+0.01}_{-0.02}$. 
\cite{Buisson_2019} found $R_{in} \sim 5.4 r_g$, consistent with a low to moderate spin. 
\citet{Bhargava_2021} analysed the characteristic frequencies of variability in \nicer\ data, modelling the power density spectrum with the relativistic precession model \citep{Stella_1998, Stella_1999, Motta_2014} and found $a = 0.799^{+0.016}_{-0.015}$.

There are also several reports of truncated discs in the literature. \citet{Dzielak_2021} suggested an $R_{in}$ in the range $\sim$ 5 - 45 $r_g$. \citet{Guan_2021} found recession of the inner radius when moving from the soft to hard states. Using fainter observations of \mx\ (less than $1$\% of the Eddington luminosity) \citet{Xu_2020} suggest up to $R_{in} \sim 300 r_g$. 

Additional methods, such as reverberation lags find equally little consensus. \citet{Kara_2019} found a decrease in the distance between the disc and corona as the source evolved through the hard state. At the same time they measured a broad Fe-K$\alpha$ line, suggesting that $R_{in}$ remained constant and close to the black hole and with the corona gradually becoming more compact. This is disputed by the modelling of \citet{DeMarco_2021}, who measured an increasing disc temperature suggesting instead that $R_{in}$ decreases.

In our spectral analysis (see Section~\ref{EPIC_pn_NuSTAR}), we use variations of our model with different spin values. We use a model with a free spin parameter, but we also make use of two subsets of this model: a low spin version ($a = 0$) and a high spin version ($a = 0.998$). We use these to better understand the effect of low or high spin on the spectral model and on the how it fits the data. These fixed-spin models are nested within the more general free-spin model, and so we could use the SDDR to compare models with different values of spin parameter.

\section{RGS spectral fitting} 
\label{RGS_section}

\subsection{The RGS model} 
\label{RGS_model}

Soft X-ray absorption by the interstellar medium (ISM) has a major effect on the shape of the spectrum below $\sim 2$ keV for many Galactic X-ray sources, including most X-ray binaries. The ISM absorption needs to be modelled simultaneously with the X-ray continuum, and it is important to use an accurate (unbiased) absorption model to avoid biasing the parameters of the emission model. 
Here, we use a similar procedure to that of \cite{Eckersall_2017}. Namely, we exploit the high resolution of the RGS data to constrain the absorption column densities of several of the species responsible for absorption in the spectrum of \mx, and use these constraints to define priors for use in subsequent modelling of the EPIC pn and \nustar\ data.

\begin{figure}
	\includegraphics[width=\columnwidth]{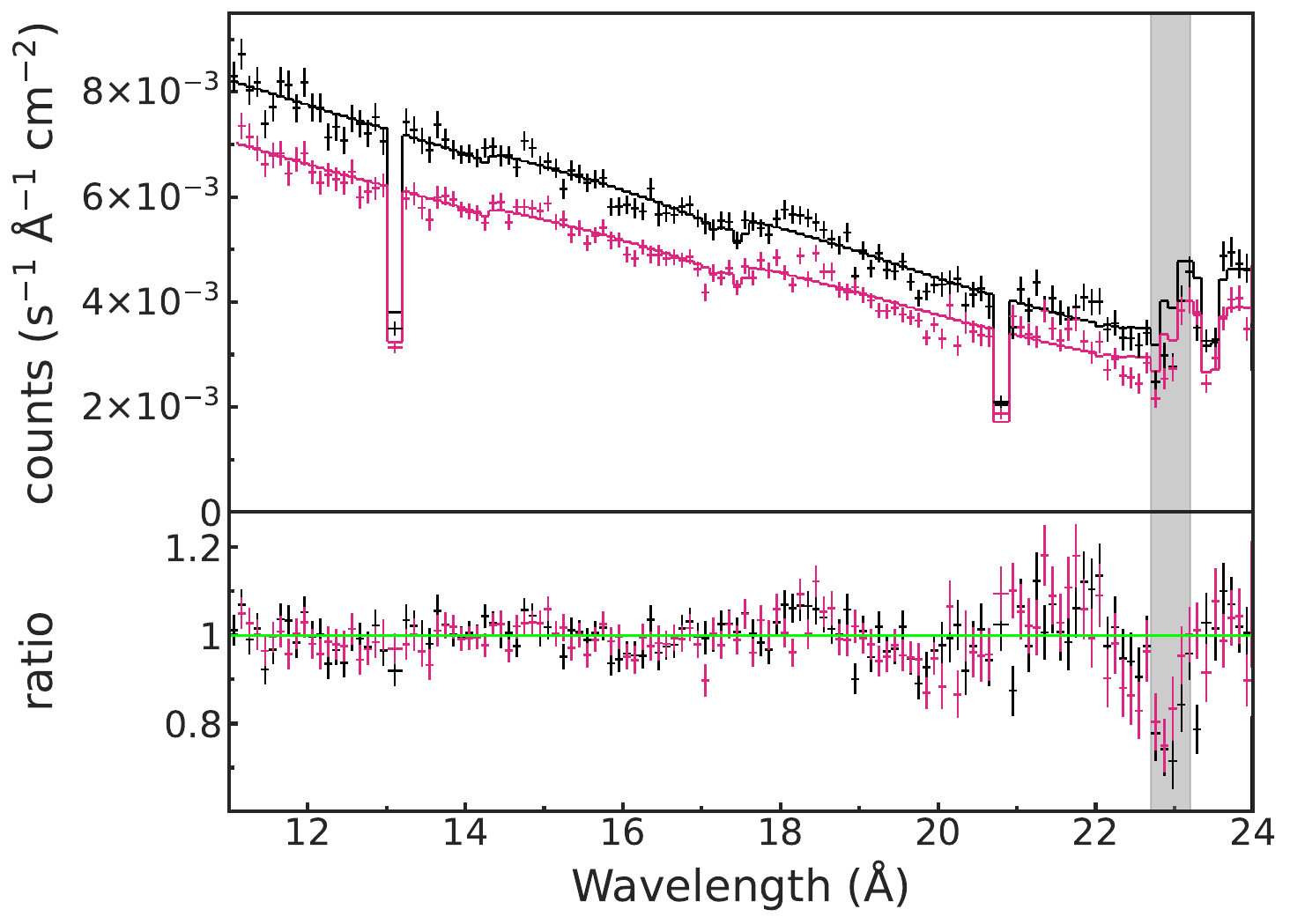}
    \caption{The RGS Spectra for \mx\ obs1 (black) and obs2 (pink), with binning such that each bin had at least 60 sigma or was grouped in sets of 10 bins. The effects of Ne, Fe and O on the spectra can be seen at $\sim$ 14, 17.5 and 23 \AA {} respectively. A shaded grey box defines the region 22.7 - 23.2 \AA{}, which was ignored due to the effects of dust.}
    \label{fig:RGS_spec}
\end{figure}

The RGS data from obs1 and obs2 had similar count rates (see table \ref{tab:OBS_TABLE}) while the count rate during obs3 was an order of magnitude lower. We therefore ignored the obs3 RGS data and modelled obs1 and obs2 simultaneously, see Fig. \ref{fig:RGS_spec}. For the simultaneous fitting, the ISM absorption parameters were tied between the two spectra, and the continuum parameters were independent to account for any changes in the source luminosity or spectrum. 

The model we apply to the RGS data is:
\[
\centerline{ {\tt ISMabs * (SIMPL * KERRBB).} }
\]
The emission over the $0.3-1.5$ keV RGS range is modelled with {\sc KERRBB} \citep{Li_2005} and {\sc SIMPL} \citep{Steiner_2009}. The former models the thermal emission from a relativistic accretion disc, the latter uses a power law to approximate inverse-Compton scattering of soft photons from the disc. Further details of these components are discussed in the next section. Together, these form an approximation to the spectrum of a typical X-ray binary. At this stage, we are not interested directly in the continuum model parameters, our focus is on inferring the parameter of the absorption model. In order to model ISM absorption features, {\sc ISMabs} \citep{Gatuzz_2015} was used as this is the most detailed available X-ray photoabsorption model for the gaseous ISM, and accounts for neutral and ionized species for a variety of elements. 

In practice, many elements whose effects are included in the model have negligible impact on the spectrum (e.g. C, Mg, Si, S, Ar and Ca).
We tied the column densities of the neutral species of these elements to that of H, using relative abundances from \cite{Wilms_2000}, and set the column density of ionised species to zero. The column density of He is fixed to $0.1$ that of H in the model, as they are highly covariant in the fitting. The column densities of H, O, Ne and Fe (including singly and doubly ionised O and Ne) were left as free parameters, as they had the largest impact on the spectrum in the observed range. 

The effects of absorption by dust in the ISM are not included in {\sc ISMabs}. We examined the potential impact of dust using the {\sc tbvarabs} model \citep{Wilms_2000} that includes fewer parameters controlling the gaseous absorption but does include absorption features from dust (grains) and molecules. We included {\sc tbvarabs}, over the {\sc ISMabs} model, to model primarily dust by setting the grain depletion fractions of all elements to $1.0$, in order to maximise the dust absorption and negate any gas absorption. We found this had a negligible effect on the quality of the fit but with more parameters, and it was not possible to separate the effects of gas from dust around the O-Edge. We therefore used only {\sc ISMabs} (with no dust) for further modelling -- i.e. assuming that all the absorption important in the $>0.8$ keV region is due to gas in the ISM -- and we ignored a small region of the RGS spectrum where the effect of dust is expected to be strongest (see Section~\ref{xmm_newton}). 

The continuum model parameters are {\it nuisance} parameters\footnote{Nuisance parameters are free parameters that do change the model predictions but we are not directly interested in their values.}. Here, we are interested in the column densities by fitting absorption features over a continuum model. The details of the continuum model are unimportant; the continuum model only needs to give a reasonable match to the data, sufficient to recover the absorption features over the continuum. 

Several parameters of the {\sc KERRBB} continuum were fixed at zero: $a$, redshift ($z$) and the ratio of the disk power produced by a torque at the inner boundary to the disk power arising from accretion ($\eta$; see Section~\ref{disco_assumptions}). The disc colour correction factor ($f_{col}$; spectral hardening factor) was fixed at $1.7$. The normalisation parameter was fixed to $1$ such that the disc luminosity is set by the accretion rate ($\dot{M}$) parameter. The effects of both self-irradiation and limb darkening were included, with the addition of both up and down scattering from {\sc SIMPL}. For the remaining four parameters of the {\sc KERRBB} continuum, we assigned very broad (diffuse) priors, despite having strong prior information about some of these (e.g. $M_{BH}$, $i$ and $D$ - see previous section). In particular, we assigned uniform priors (over a wide range) on all free {\sc KERRBB} and {\sc SIMPL} parameters except for $\dot{M}$ which was assigned a uniform prior in its logarithm, i.e. $p(x)\propto 1/x$. The rationale for this is that the objective of this exercises is to obtain good inferences about only the absorption column densities in a manner that is robust to assumptions about the details of the continuum. 

\subsection{Parameter inference}

\begin{figure}
	\includegraphics[width=\columnwidth]{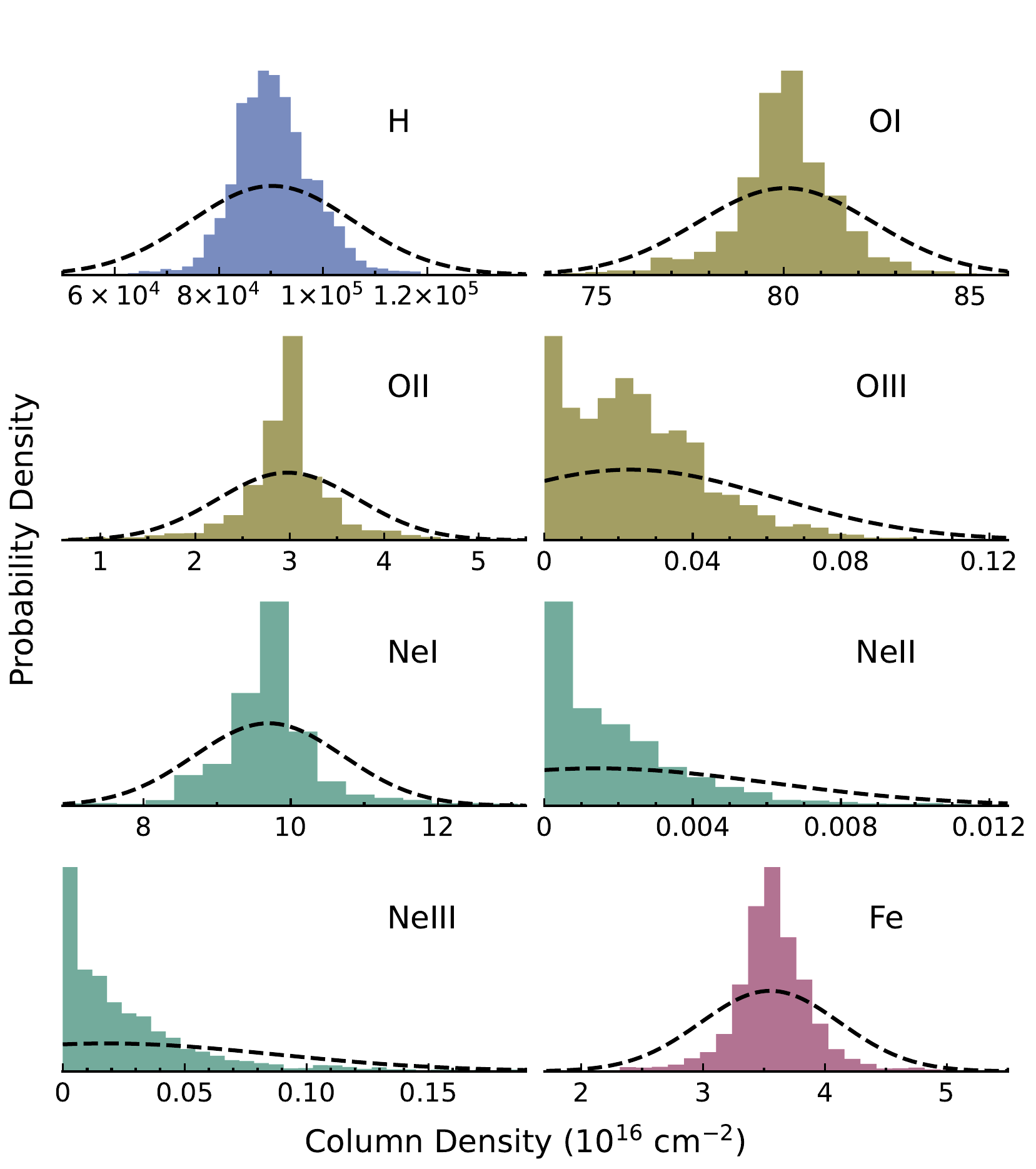}
    \caption{Histograms of the posterior samples for each column density from the RGS fitting of \xmm\ data. Each plot is labelled referring to which column density is displayed. A black dotted line has been added to represent the Gaussian priors adopted (with twice the width of the posterior samples) for the fitting of EPIC pn and \nustar\ data.}
    \label{fig:RGS_priors}
\end{figure}

Having defined the model and prepared the spectra, we ran MCMC to sample from the posterior (see Appendix \ref{MCMC_Appendix} for details). The resulting posteriors are shown in Fig. \ref{fig:RGS_priors}, and a summary of each of the key parameters is given in Table \ref{tab:RGS_results}. We used the posterior median and $16^{th}$ and $84^{th}$ centiles, to describe the centre and a `$1$-sigma' ($68$\%) interval around this.

Our $N_H$ estimate of $0.90^{+0.08}_{-0.06} \times 10^{21}$ cm$^{-2}$ is below that estimated from $21$ cm surveys. The LAB survey gives a value of $1.4 \times 10^{21}$ cm$^{-2}$ \citep{Bharali_2019, Kalberla_2005} or $2.2 \times 10^{21}$ cm$^{-2}$ when corrected for $H_2$ \citep{Willingale_2013}. The $21$ cm maps trace the total H column and so will systematically overestimate the X-ray absorption towards relatively nearby objects like \mx\ at $D \approx 3$ kpc. 

Our inferred $N_H$ value compares well to that from previous X-ray spectral studies, most of which use the \textsc{tbabs} or \textsc{tbnew} models. 
Previous estimates span that range $\sim (0.5 - 2.0) \times 10^{21}$ cm$^{-2}$ \citep{Fabian_2020, Draghis_2023, Espinasse_2020}. Our column density lies comfortably in this range, with close agreement to values from \cite{Shidatsu_2018}, \cite{Xu_2020} and some values from \cite{Draghis_2023}. None of these previous studies fitted for the relative abundances of O, Ne etc. in the ISM, and so we cannot compare other column densities. 
We note that our column densities for neutral O, Ne and Fe are slightly higher (typically by a factor of $1.2-1.8$) than expected based on the relative abundances of \cite{Wilms_2000} . 

\begin{table}
	\centering
	\caption{Column densities given by the MCMC analysis of RGS spectra from \mx. As almost all errors were asymmetric, both the upper and lower error is presented. The error displays the range between the 16$^{th}$ centile and the median and the median and the 84$^{th}$ centile respectively.}
	\label{tab:RGS_results}
	\begin{tabular}{rcc}
		\hline 
		\hline 	
		Ionic Species & Column Density & Error \\
		\hline
		H (10$^{22}$ cm$^{-2}$)& 0.090 & (+0.008)(-0.006)\\
		OI (10$^{16}$ cm$^{-2}$)& 80.056 & (+1.175)(-1.101)\\
		OII (10$^{16}$ cm$^{-2}$)& 2.977 & (+0.332)(-0.376)\\
		OIII (10$^{16}$ cm$^{-2}$)& 0.023 & (+0.020)(-0.017)\\
		NeI (10$^{16}$ cm$^{-2}$)& 9.700 & (+0.466)(-0.506)\\
		NeII (10$^{16}$ cm$^{-2}$)& 0.001 & (+0.002)(-0.001)\\
		NeIII (10$^{16}$ cm$^{-2}$)& 0.018 & (+0.035)(-0.015)\\
		Fe (10$^{16}$ cm$^{-2}$) & 3.553 & (+0.285)(-0.229)\\
		\hline 	
		\hline 	

	\end{tabular}
\end{table}

The column density posteriors discussed above were computed using MCMC. We also tried the NS method on the RGS data. The resulting posteriors were slightly broader than those from MCMC -- in some cases -- but we found only marginal differences in the final results when these were carried through to the analysis of EPIC and \nustar\ spectra discussed below.

\section{The vanilla Model}
\label{Vanilla_model}

\subsection{Outline}
\label{Vanilla_model_model}

Our {\it vanilla} model for the broad-band ($0.8-70$ keV) X-ray spectrum of \mx\ is:
\[
\centerline{\sc ISMabs * (SIMPL * KERRBB + relxillCp).}
\]
We chose model these components over alternative possibilities in order to be as self consistent as possible, while paying attention to computational speed and also allowing for the relaxation of some common assumptions intrinsic to alternative components. The first three components of the {\it vanilla} model are largely the same as stated in Section \ref{RGS_model}. 

\subsection{Model components}
\label{Vanilla_model_components}

The \textsc{ISMabs} component (discussed above) accounts for absorption in the ISM, and here has eight free parameters. We use our posterior samples from the RGS analysis to define priors on these. We approximate the posteriors of the eight column densities as Gaussian priors, see Fig. \ref{fig:RGS_priors}. All the posteriors used were slightly asymmetric (see Table \ref{tab:RGS_results}). For the standard deviation of our Gaussian prior on each of these parameters, we used double the larger of the two distances between the median and the $(16, 84)$ centiles. By making these priors wider than the posteriors from the previous section, we allow for the possibility of some systematic errors (e.g. in the calibration of the RGS or the treatment of the background) and reduce the sensitivity of our final results to the fine details of the RGS analysis.

The {\sc KERRBB} component represents the thermal emission from the accretion disc, assuming it is optically thick, geometrically thin, and extends into the ISCO. The model has $6$ free parameters: $f_{col}$, $a$, $i$, $M_{BH}$, $D_{BH}$ and $\dot{M}$. Here, we treat $f_{col}$ as a free parameter, as it remains unclear what the physically realistic value should be -- see e.g. \cite{Davis_2019} and \cite{Salvesen_2013}. The redshift ($z$) was fixed at zero and we fixed $\eta=0$ (meaning zero torque at the inner edge of the disc). 

The {\sc SIMPL} component has two free parameters ($f_{scat}$ and $\Gamma$). We allow for up- and down-scattering. This produces a power law spectrum of photon index $\Gamma > 1$ and conserves photon number in the sense that all the photons in the hard X-ray power law are subtracted from {\sc KERRBB}. Specifically, a fraction ($0 \le f_{scat} \le 1$) of photons from the {\sc KERRBB} component are re-distributed to give the power law. This approximates the properties of a hard X-ray continuum produced by inverse Compton scattering of seed photons from the optically thick accretion flow: the hard X-ray luminosity is linked to that of the seed photon source (the thermal disc) and does not contribute strongly below the peak energy of the seed photon spectrum. The commonly-used alternative, namely adding a simple power law component, does not conserve photon number in this way and results in 'flux stealing' between the thermal (seed) and non-thermal (Compton scattered) components. A simple power law, when extrapolated to low energies, overestimates the soft X-ray contribution compared to a more realistic model, and leads to systematically underestimated seed photon flux. See \cite{Salvesen_2013} for discussion of this point. We tried the more complex {\sc ThComp} model \citep{Zdziarski_2020} as an alternative, but found it that in practice it produced in worse fits with no computational advantage.

The model component \textsc{relxillCp} was used to model the `X-ray reflection' spectrum from \mx. The \textsc{relxill} \citep{Dauser_2014, Garcia_2014} suite represents the most accurate and self-contained models of the reflection spectrum. These models are themselves comprised of two components: \textsc{xillver} \citep{Garcia_2010, Garcia_2011, Garcia_2013}, which models the local reflection spectrum and \textsc{relline} \citep{Dauser_2010, Dauser_2013} which uses ray tracing to apply relativistic distortions to the rest-frame \textsc{xillver} spectrum. There are a variety of different flavours of this model, including  \textsc{relxillCp}, which differs from the basic \textsc{relxill} in its use of an {\sc Nthcomp} (\cite{Zdziarski_1996, Zycki_1999} Comptonisation continuum in place of a high-energy cutoff powerlaw. 

An additional advantage -- and why we judge this model as more suitable -- is that \textsc{relxillCp} allows for higher density of the reflecting medium ($n_e$ up to $10^{20}$ cm$^{-3}$; by comparison, the basic \textsc{relxill} model fixes $n_e$ at $10^{15}$ cm$^{-3}$). 
We might expect the electron density in the inner disc to be higher then this (by $1-2$ orders of magnitude). The reflected continuum spectrum depends on this density, and using models with a lower density can bias other {\sc relxillCp} parameters. It is also the case that the atomic physics of higher density X-ray reflection is less well understood. See 
\citet{Garcia_2016} and \citet{Tomsick_2018} for discussion of these points. For this reason, $A_{Fe}$ and $\Gamma_R$ are considered as nuisance parameters. \cite{Liu_2023} further supports the need for higher disc densities. They used \nustar\ and \swift\ observations of six black hole X-ray binaries, estimating that disc densities in excess of $10^{15}$ cm$^{-3}$ were required in most cases.

There are a total of six free parameters in \textsc{relxillCp}: two emissivity indices ($I_1$ \& $I_2$) representing the X-ray illumination of the reflector, the power-law index of the incident spectrum ($\Gamma_R$), the ionisation of the reflector ($\log(\xi)$), its iron abundance ($A_{Fe}$), and a normalisation parameter ($norm_R$). The inclination of the reflector is tied to the inclination of the {\sc KERRBB} component ($i$, above), as is the black hole spin parameter, and the redshift is fixed at $z=0$. We fix $R_{in}=R_{ISCO}$ and $R_{out} = 1000$ $r_g$ (the maximum allowed by the model). 

$\Gamma_R$, describes the photon index of the hard X-ray power law spectrum that illuminates the reflector and is assumed to originate in a corona above the disc. In the usual picture, $\Gamma_R$ would be coupled with $\Gamma_S$ from \textsc{SIMPL} as the standard assumption is that the reflector is illuminated by the same spectrum as we observe directly in hard X-rays. In our model, we allow these two photon indices to vary independently. This allows the model to account for the possibility that the coronal emission is anisotropic -- where the spectrum seen by the reflector is different to that seen directly by a distant observer, as may be expected if the corona is coincident with the base of a jet \citep{Markoff_2004}. This also allows the model to `absorb' some possible biases caused by incompleteness of the microphysics of the reflection spectrum (see above discussion of disc density). In essence, we treat $\Delta \Gamma = \Gamma_S - \Gamma_R$ as a nuisance parameter. The illuminating spectrum provided by the {\sc Nthcomp} Comptonisation model is well approximated by a power law up to energies $\sim kT_e$, the electron temperature, which we fix at the upper bound of $400$ keV. Relaxing this assumption made no difference to the fits and made the fitting/sampling process much more challenging.

The emissivity indices (together with $R_{in}$, $R_{out}$ and $i$) control the shapes of the relativistic distortion on the reflection spectrum. The emissivity over the reflector is modelled as a broken power law from $R_{in}$ to $R_{out}$, with index $I_1$ inside some fixed break radius (R$_{br}$), and index $I_2$ outside of this. We fixed $R_{br} = 18$ $r_g$ here (see Section~\ref{disco_assumptions}). The exact choice of $R_{br}$ is not important. The motivation behind this is to allow for some differences in the emissivity between the `inner' (highly relativistic) parts of the disc close to ISCO, and the `outer' disc beyond this. Somewhat arbitrarily, we chose a value for $R_{br}$ that corresponds to $3R_{ISCO}$ for $a=0$ to distinguish `inner' and `outer' regions of the disc.

The final parameter, $norm_R$, sets the absolute normalisation of the reflected spectrum. The parameter that sets the relative strength of the reflection to incident spectrum is fixed to $R=-1$. This means the reflection strength is controlled only by $norm_R$ and \textsc{relxillCp} adds only reflection, the incident power law is not included.

\begin{table*}
	\centering
	\caption{Parameter  values  used  for  each component  of  the  vanilla  model. Due to the high number of parameters in \textsc{ISMabs}, only those that were free are shown. The full range of permitted values for each parameter are given in their description. Where '$*$' is used, it indicates when the parameter is fixed at the stated value. Where priors are labelled 'Tied', they are tied to their respective counterpart in \textsc{KERRBB}. Priors denoted by a '$T$' were truncated for NS, to prevent values from being chosen outside of the hard limit of the model. 
    The limits of the {\sc relxillCp} normalisation were adjusted for each observation depending on the brightness.}
	\label{tab:Params}
	\begin{tabular}{l l l  c c}
 	\hline
 	\hline
	Model & Parameter & Description & Prior &  Units\\

 	\hline
 	\hline
	\textsc{ISMabs} & $N_H$    &H column density [0, 10$^6$]&  Gaussian&10$^{22}$ cm$^{-2}$ \\
	   &$N_{OI}$    &OI column density [0, 10$^6$]&  Gaussian&10$^{16}$ cm$^{-2}$\\
	   &$N_{OII}$    &OII column density [0, 10$^6$]&  Gaussian&10$^{16}$ cm$^{-2}$\\
	   &$N_{OIII}$    &OIII column density [0, 10$^6$]&  Gaussian$^T$&10$^{16}$ cm$^{-2}$\\
	   &$N_{NeI}$    &NeI column density [0, 10$^6$]&  Gaussian&10$^{16}$ cm$^{-2}$ \\
	   &$N_{NeII}$    &NeII column density [0, 10$^6$]&  Gaussian$^T$&10$^{16}$ cm$^{-2}$\\
	   &$N_{NeIII}$ &NeIII column density [0, 10$^6$]&  Gaussian$^T$&10$^{16}$ cm$^{-2}$ \\
	   &$N_{Fe}$    &Fe column density [0, 10$^6$]&  Gaussian&10$^{16}$ cm$^{-2}$\\
       \hline
       
    	\textsc{SIMPL} & $\Gamma_{S}$    &Coronal power-law index [1, 5]&  Uniform& --\\
          &$f_{scat}$    &Scattering fraction [0, 1]&   Uniform& --\\
 		\hline
 		
 		\textsc{KERRBB} & $\eta$    &Ratio of disc power  $[0, 1]$ &   $0$*& --\\
    	&$a$    &Black hole spin $[-0.998, 0.988]$&   Uniform / 0* / 0.998*&--\\
   		&$i$    &Inclination $[3, 85]$ &   Gaussian& degrees\\
   		&$M_{BH}$    &Mass of black hole $[0, 50]$ &   Shifted log-normal& \Msun\\
   		&$\dot{M}$    & Effective mass accretion rate [$10^{-6}\dot{M}_{Edd}$, $\dot{M}_{Edd}$] &  Jeffreys &10$^{18}$ g s$^{-1}$\\
    	&$D_{BH}$    &Distance of system $[0, 30]$ &   Gamma$^T$& kpc\\
    	&$f_{col}$    &Spectral hardening factor $[1, 10]$&  Gamma$^T$& --\\
 		\hline
 		\textsc{relxillCp} &$i$    &Inclination $[3, 85]$ &   Tied& degrees\\
 		&$a$    &Black hole spin $[-0.998, 0.988]$ &   Tied & --\\
	   &$R_{in}$    & Accretion disc inner radius $[1, 100]$&   1*& $R_{ISCO}$\\
	   &$R_{out}$    & Accretion disc outer radius $[1, 1000]$&   1000* &$r_{g}$\\ 	
	   &$R_{br}$    & Emissivity break radius $[1,1000]$ &   18* &r$_{g}$\\
 	   & $I_1$    & Emissivity index ($R_{in}$ - $R_{br}$) $[0, 10]$&   Gamma$^T$& --\\
	   &$I_2$    & Emissivity index (R$_{br}$ - R$_{out}$) $[0, 10]$&   Gamma$^T$& --\\
	   &$\Gamma_{R}$    &Power law index of incident spectrum $[1, 3]$ &   Uniform& --\\
	   &$\log(\xi)$    &Ionisation of accretion disc [0, 4.7] &   Uniform& --\\
	   &$\log(N)$    &Logarithmic accretion disk density $[15, 20]$ &   20* & log(cm$^{-3}$) \\
	   &$A_{Fe}$    &Iron abundance $[0.5, 10]$ &   Uniform& --\\
	   &$kT_e$    &Electron temperature in the corona [5, 400] &   400* & keV\\
	   &$R$    &Reflection fraction [-1000, 1000] &   -1*& --\\
	   &$norm_R$    &Normalisation [see caption]&   Jeffreys& --\\
 		\hline
 		\textsc{constant} & $C$  &A constant scaling $[0.5, 3]$ &   Uniform & --\\	
 	    \hline
 	    \hline

	\end{tabular}
\end{table*}

\subsection{Priors}
\label{priors}

The model has many parameters and we assign priors to each of them. For the {\sc ISMabs} column densities, the priors come from analysis of the RGS data (Section~\ref{RGS_section}), for the system parameters ($M_{BH}$, $i$, $D$), these come from the best multiwavelength studies of the binary system (Section~\ref{MAXI_J1820}). Here we discuss the assignment of priors to the remaining parameters of the model. We assign `informative priors' to those parameters where there is a good understanding of their expected or typical values based on previous observations or strong theoretical arguments. For the remaining free parameters we assign `weakly informative priors', i.e. priors that are relatively equivocal about the parameter value over a wide range. 

For the colour correction factor $f_{col}$, most theoretical work suggests a value over the range $1.4 - 2.0$ \citep{Davis_2019}, with $1.7$ being the canonical value following \citet{Shimura_1995}. However, \citet{Salvesen_2013} suggests a larger range of plausible values up to $5.0$. In order to encode this, We assigned Gamma density to  $f_{col}$ with hyperparameters chosen such that the prior mode is $f_{col} = 1.7$ and the probability density is high over the $1.4 - 2.0$ range, and moderate up to $5.0$. Above $5.0$ the prior has a total probability mass of $\sim 0.006$. 

For the two emissivity parameters - $I_1$ and $I_2$ - we also used Gamma priors. The broken power law emissivity is necessarily an approximation to a more complicated situation. General relativistic ray-tracing simulations of different geometries typically give emissivity profiles that are reasonably approximated by broken power laws with indices of $\sim 3$ at large radii and steeper at smaller radii \citep{Wilkins_Fabian_2012}. We employ the same prior for both indices, with hyperparameters $\alpha$ and $\beta$ chosen such that the prior density peaks at $3.0$ ($\alpha = 13.23$ and $\beta = 4.08$) and large values ($>5$) have low probability. 

We assigned independent Jeffreys priors ($p(x) \propto 1/x$) on two `scale parameters': $\dot{M}$ and $norm_R$. These are invariant to a multiplicative rescaling of the parameter; see \cite{Practical_Bayesian_Inference_Book} for discussion of priors for scale parameters. In its simplest form, this prior is improper and diverges as the parameter tends to zero. These can cause issues with improper posteriors, making model checking difficult, and difficulty sampling from the posterior, and so we impose positive-valued lower and upper limits on these parameters.

For the upper limit of $\dot{M}$ we use the Eddington mass accretion rate ($\dot{M}_{Edd} \approx 23.3 \times 10^{18}$ g s$^{-1}$, based upon the $90$th centile of the mass distribution; see above). \citet{Shaw_2021} tracked \mx\ into quiescence, down to an Eddington fraction of $\dot{M}/\dot{M}_{Edd} \sim 4 \times 10^{-7}$. The present observations were made during re-brightening episodes, and so we assume a lower limit of $\dot{M}/\dot{M}_{Edd} \approx 10^{-6}$ ($\dot{M} \approx 18.6 \times 10^{12}$ g s$^{-1}$, based upon the $10$th centile of the mass distribution). For the $norm_R$ parameter, which defines the reflected flux, we use as an upper limit a normalisation that would greatly exceed the total observed flux, and as a lower limit the normalisation that would give only $\sim 1$ reflected photon in the spectrum (i.e. below the limit to which that data are sensitive). Specifically, we compared each observation to a model comprising only {\sc ISMabs * relxillCp} but with reflection strength $R = 0$ such that {\sc relxillCp} contributes only a power law. Setting $\Gamma_R = 1.9$, $kTe = 50$ keV and using ISM parameters from Section \ref{RGS_section} we find values of $norm_R$ of $\sim 1.7 \times 10^{-3}$, $1.4 \times 10^{-3}$ and $1.7 \times 10^{-4}$ for obs1, obs2 and obs3 respectively give fluxes similar to those observed. We set the upper limit on $norm_R$ as $100$ larger than this, and set the lower limit to $10^{-6}$ of this. 

For $a$, we allow the parameter to vary across all possible values, with a uniform prior over the range $[-0.998, +0.998]$. Finally, we assign uniform priors over a range $(0.5,3)$ to each of the normalising constants used to scale between the \xmm\ EPIC pn data and both \nustar\ FPM-A and FPM-B spectra. 

A full list of all {\it vanilla} model components, parameters and choice of values or priors can be seen in Table \ref{tab:Params}. 

For some parameters, the priors defined above extend (with non-zero probability) outside the range allowed by the model. This would cause issues using NS, as new points are drawn from the priors. Some parameters with relatively narrow priors (e.g. $M_{BH}$ or $i$) caused no problems. For parameters where this did cause problems with sampling (see the next section), we truncated the prior at the limits of the model. Further, when fitting to \nustar\ data, we enforced $\Gamma_R \le 3$ to avoid problems with the computation\footnote{Without this constraint, the sampler would occasionally propose values of $\Gamma_R \ge 3$ that often caused the code to terminate prematurely. We practically eliminated this computational issue by restricted the range of $\Gamma_R$.}.

\section{EPIC and \nustar\ SPECTRAL FITTING} 
\label{EPIC_pn_NuSTAR}

We are now ready to model the EPIC pn and \nustar\ data; we have a spectral model defined, with a likelihood function and a prior density. Together, these define a posterior density (for the model and its parameters, given the data). We now proceed to sample from the posterior using both MCMC and NS methods. To recap: we use a Gaussian likelihood ($\chi^2$ as the minus log likelihood function) for MCMC sampling, and a Poisson function ($W$-stat as the log likelihood) for NS, and the prior density is the product of independent priors on each parameter. Details of how each of the two samplers were set up are given in the Appendix. 

The output of each of the Monte Carlo methods (MCMC and NS) is a large set of parameter values; unlike model optimisation (e.g. min $\chi^2$ fitting) there is no single best set of parameters. We summarise the results in various ways. One common approach to visualising a posterior density is a corner plot, showing the marginal posterior density estimates for each pair of parameters.
We have used such plots, but for publication we present a subset of the one- and two-parameter marginal distributions.
Tables \ref{tab:Param_values_MCMC} and \ref{tab:Param_values_NS} show the median and `one sigma' ranges for marginal distributions of each parameter, for each variation of the model, for each of the three observations. Fig.~\ref{fig:violinpl} illustrates the marginal posteriors for a subset of the parameters based on the obs2 data.

For obs1 and obs2, we find reasonable agreement between the posterior estimates from each of the two sampling methods (MCMC and NS) for most parameters. 

For obs3, the two sampling methods showed obvious disagreement in many of the parameters despite the key diagnostics indicating that each method had converged. For this reason, and the stronger disagreement between the instruments in obs3, we focus our attention on obs1 and obs2 and return to obs3 later. 

The marginal posteriors for some parameters -- e.g. distance and $\dot{M}$ -- are slightly narrower from the NS samples. However the posterior mode is approximately the same for both methods. The only major difference is in the $f_{scat}$ posterior samples: the mode of the samples from MCMC and from NS differ by $\sim 15$\%; two estimates of the posterior are shifted relative to each other.

In order to visualise the model and the data-model residuals, we repeated this for spin values fixed at $a = 0$ and then $a = 0.998$, equivalent to applying a different delta function priors on $a$. We then used the parameters at the posterior mode, i.e. optimising the posterior (including contributions from the fit statistic and the priors, not simply minimising $\chi^2$). The number of parameters is relatively large, and this can make it challenging to optimise the model. From the histograms of the MCMC and NS samples we could approximate the location of the mode, and then use interactive fitting to obtain a better approximation. (We used the {\tt steppar} function in {\it XSPEC} to perturb each parameter in turn and check for local minima in each dimension of the parameter space.) 

\begin{figure}
    \includegraphics[width=\columnwidth]{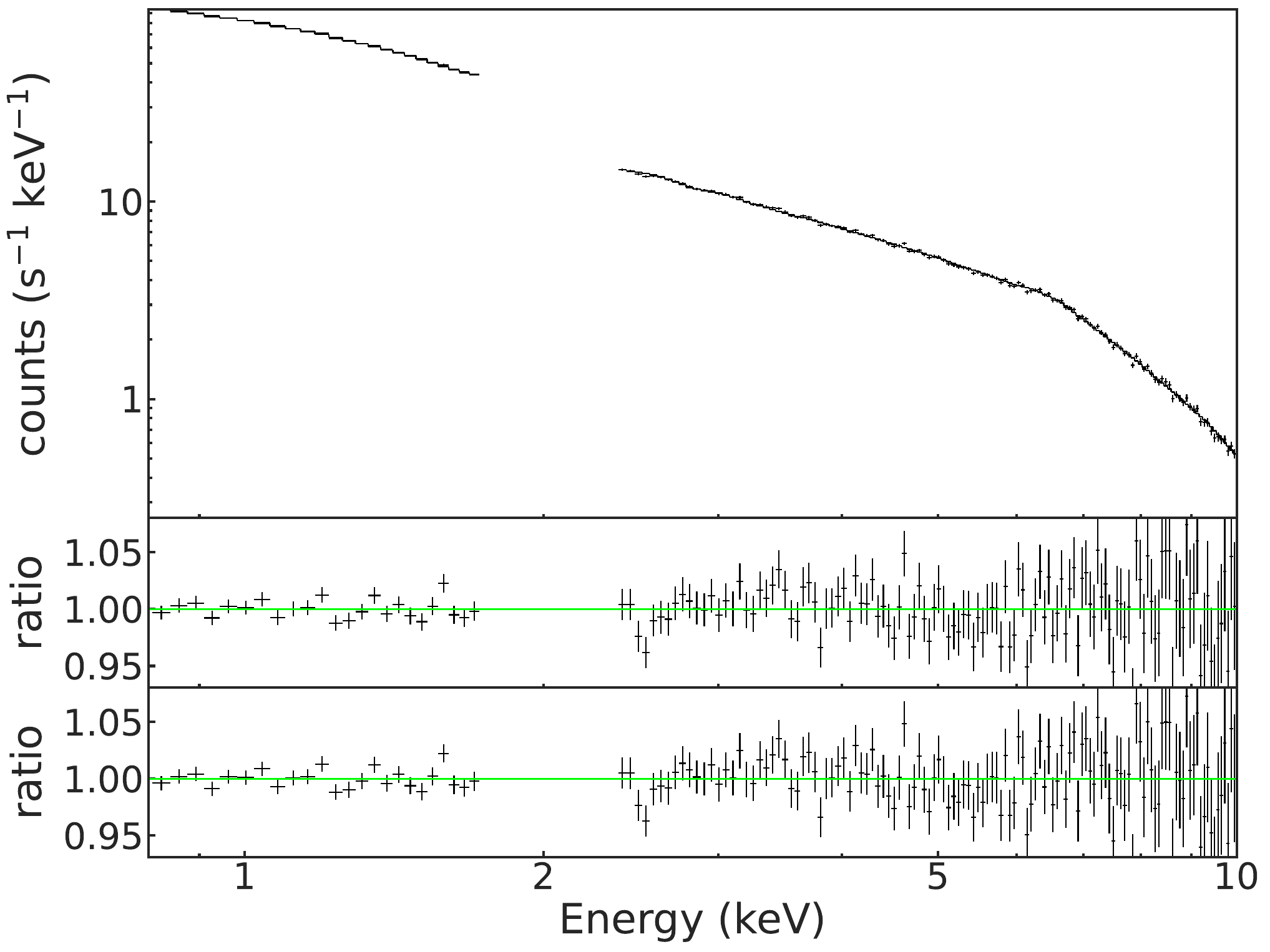}
    \caption{Top panel: The EPIC pn data for obs1 together with the low spin model. 
             Middle panel: data/model ratio for the low spin model evaluated at the posterior mode. 
             Bottom panel: data/model ratio for the high spin model.}
    \label{fig:RESIDUALS_rev3531}
\end{figure}

\begin{figure}
	\includegraphics[width=\columnwidth]{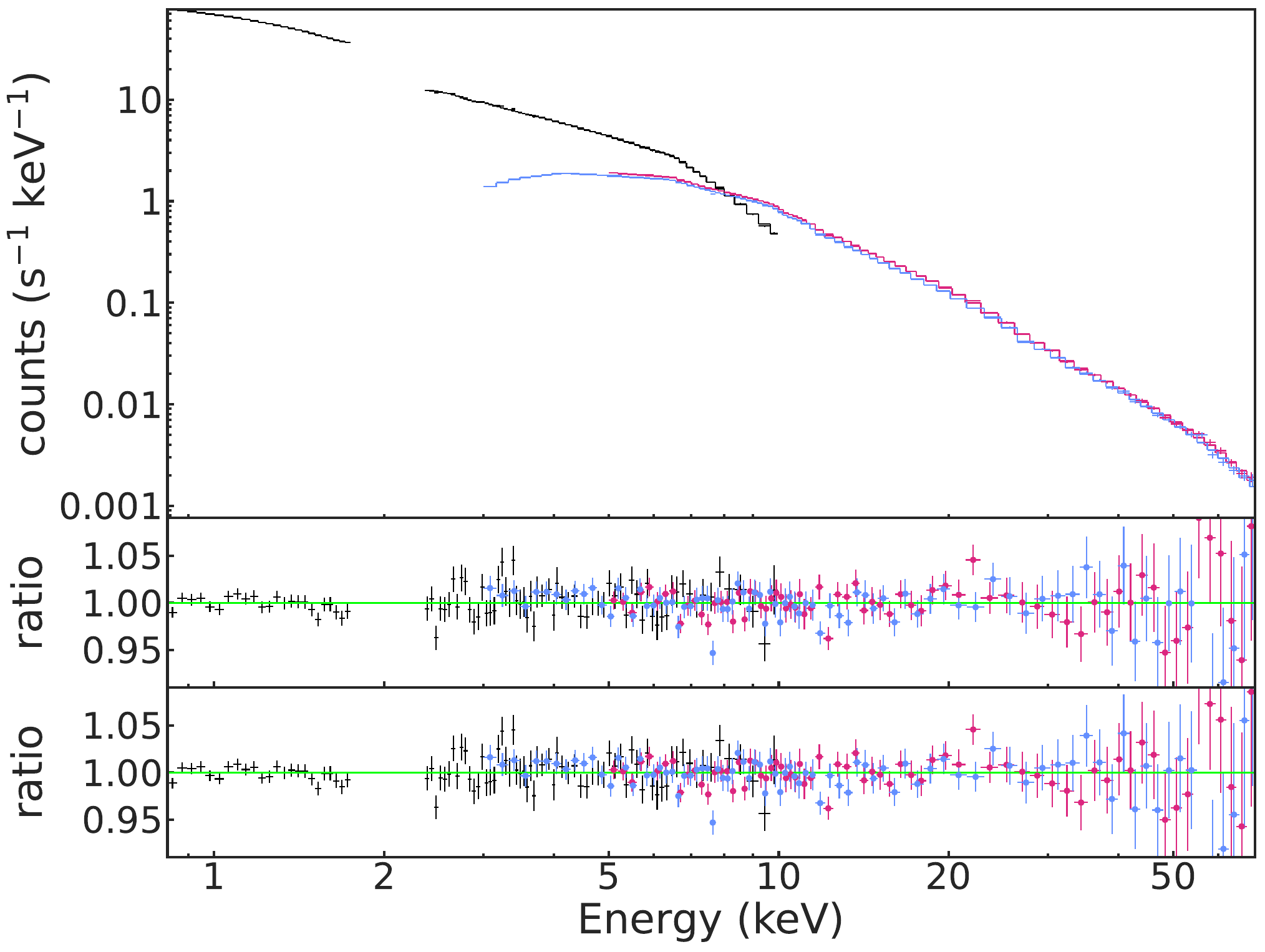}
    \caption{Same as Figure \ref{fig:RESIDUALS_rev3531} but for the obs2, showing EPIC pn (blue) and \nustar\ data (FPM-A in blue, FPM-B in pink). For the purpose of illustration only, the data have been rebinned. Both models appear to fit well and the EPIC pn and \nustar\ data are broadly consistent where they overlap.}
    \label{fig:RESIDUALS_rev3533}
\end{figure}

\begin{figure}
	\includegraphics[width=\columnwidth]{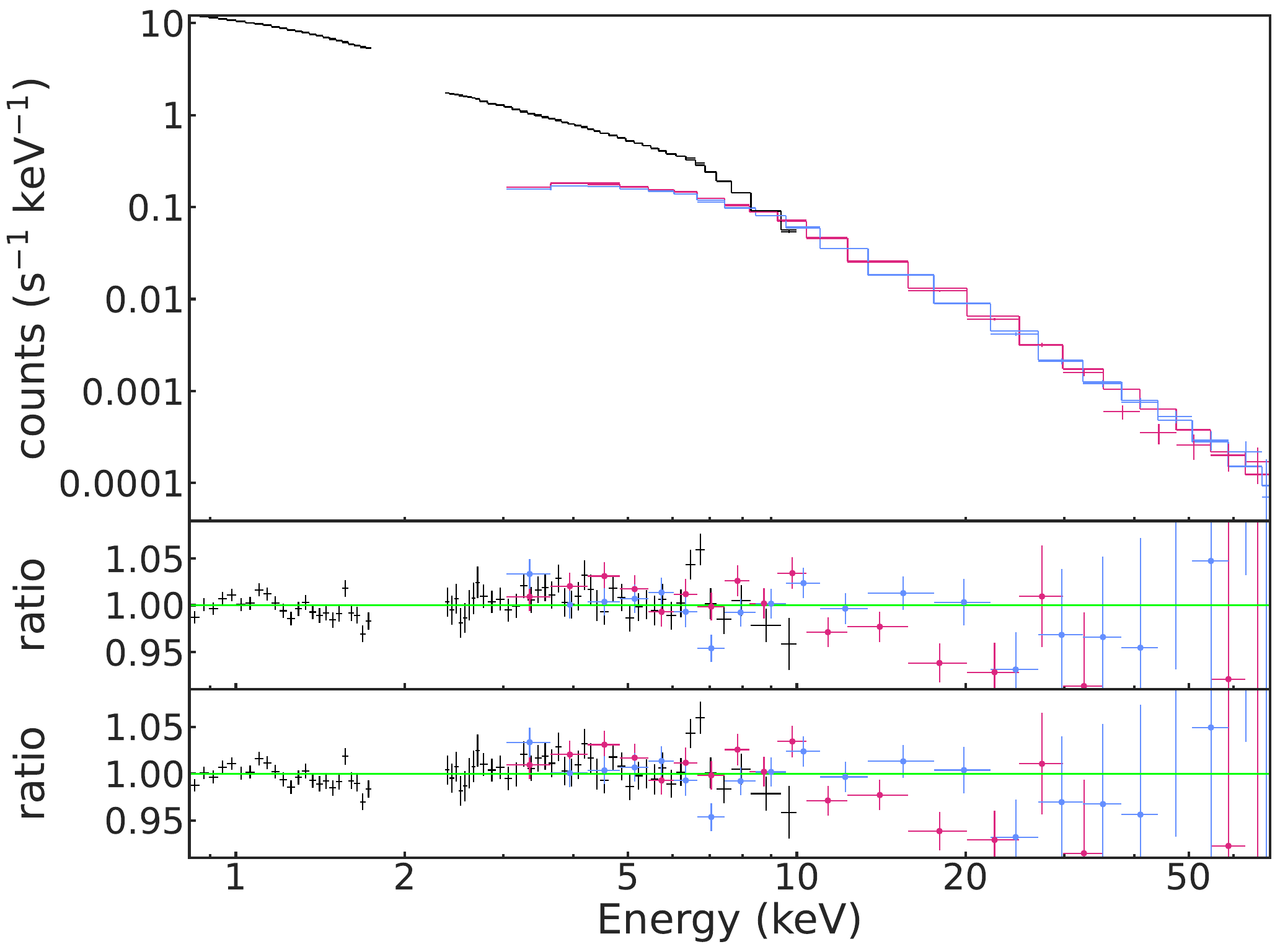}
    \caption{Same as Figure \ref{fig:RESIDUALS_rev3533} but for the obs3, showing EPIC pn and \nustar\ data. The EPIC pn and \nustar\ residuals differ noticeably around $\sim 7$ keV.}
    \label{fig:RESIDUALS_rev3623}
\end{figure}

\begin{figure}
	\includegraphics[width=\columnwidth]{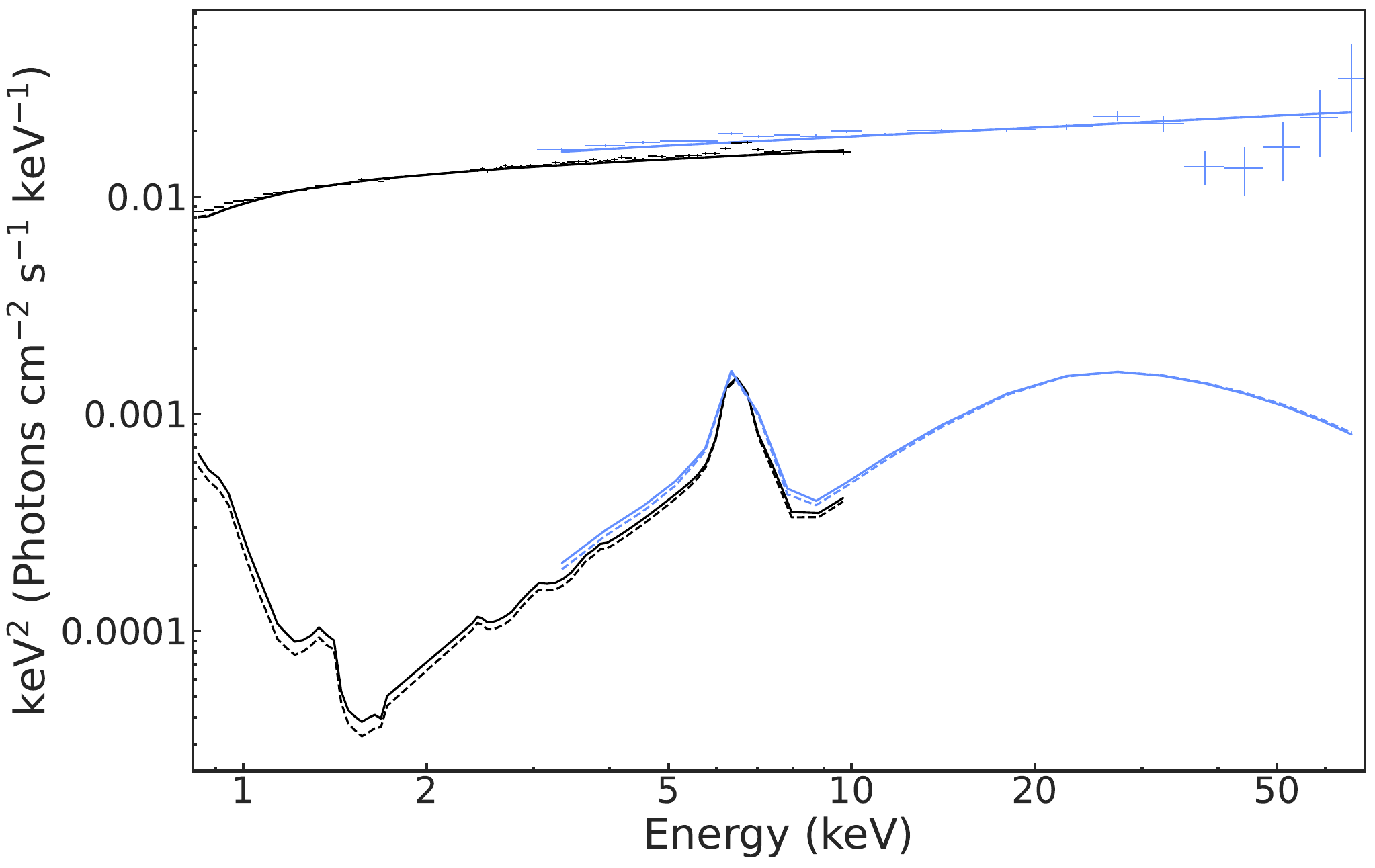}
    \caption{The unfolded spectra of obs3. The low and high spin models are shown by the dotted and solid lines, respectively. EPIC pn data is plotted in black, and \nustar\ FPM-A data in blue. The largest differences between low and high spin models occur below $7$ keV. (For clarity, FPM-B data was not plotted, but follows the FPM-A data.)}
    \label{fig:eeuf_rev3623}
\end{figure}

\begin{figure*}
	\includegraphics[width=16cm]{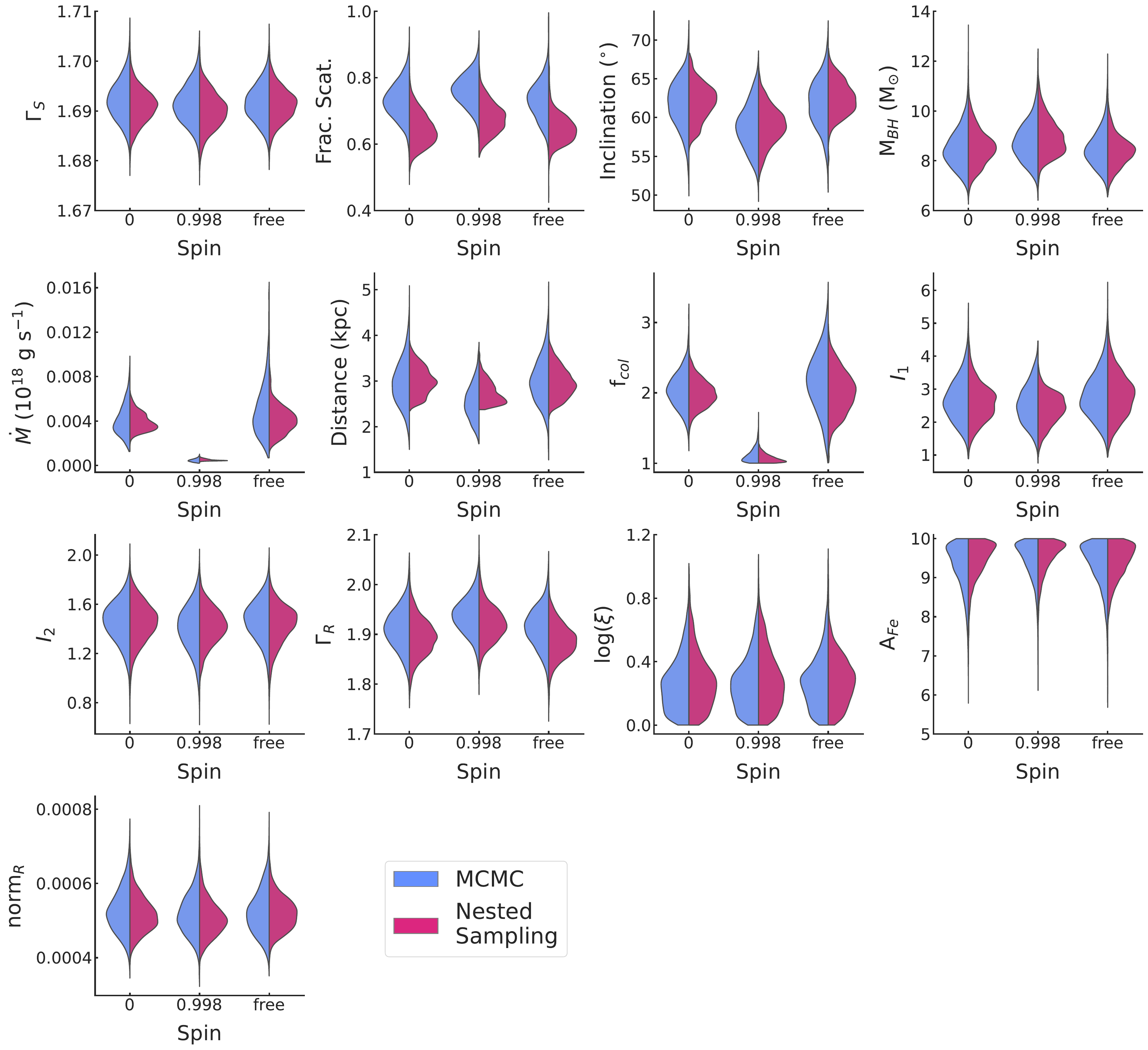}
    \caption{Smoothed posterior distributions of the {\sc SIMPL}, {\sc KERRBB} and {\sc relxillCp} model components for the obs2 data. Within each panel, we show distributions for each of the three spin models (low spin, high spin, free spin), and the left (blue) and right (pink) posteriors in each plot represent the results from MCMC and NS, respectively. }

    \label{fig:violinpl}
\end{figure*}

Following this procedure we calculated the fit statistic and data-model residuals at the posterior mode for both the high-spin and low-spin models. 
Figs. \ref{fig:RESIDUALS_rev3531}, \ref{fig:RESIDUALS_rev3533} and \ref{fig:RESIDUALS_rev3623} show the resulting models compared to the data. The $\chi^2$ fit statistics (and the corresponding log prior contributions) are given in Table \ref{tab:fit_stats}. The high and low spin models provide similar quality fits to the data -- as illustrated in Fig.~\ref{fig:eeuf_rev3623} the two fitted models give very similar predictions for the data.

\begin{table}
\caption{Summary of the fit statistics evaluated at posterior mode for each of the three observations, and for the low- and high-spin models.}
\label{tab:fit_stats}
\centering
\begin{tabular}{lllll}
\hline
\hline
obs & model & $\chi^2$ & $-2\ln(\text{prior})$ & dof \\
\hline
obs1 & $a=0$     & $131.3$  & $-26.3$  & $129$     \\
        & $a=0.998$ & $129.5$  & $-29.3$  & $129$     \\
obs2 & $a=0$     & $535.4$  & $-23.6$  & $560$     \\
        & $a=0.998$ & $533.8$  & $-24.7$  & $560$     \\
obs3 & $a=0$     & $585.8$  & $-26.9$  & $501$     \\
        & $a=0.998$ & $582.5$  & $-27.8$  & $501$     \\
\hline
\hline
\end{tabular}
\end{table}

\subsection{System parameters}

The posterior distributions of the binary system parameters ($M_{BH}$, $D$, $i$) are similar to the priors for those parameters. This happens when the data are not informative about those parameters (the likelihood is much broader than the prior and so the posterior shape is dominated by the prior). Nevertheless, these parameters are important for the modelling.

We used informative priors -- based on multi-wavelength studies of the binary system -- which ensure that the model as a whole is consistent with what we know about the system. 
Many of the parameters are covariant in the fitting which can lead to degeneracies in the parameters if few of the parameters are constrained by the data. The prior information about the system parameters, in combination with the model, helps to constrain other parameters. E.g. $D$ and $\dot{M}$ (and to a lesser extent $M_{BH}$ and $i$) tend to be well correlated in the fitting if left as free parameters with broad priors, but the informative priors on $D$, $M_{BH}$ and $i$ lead to more informative posteriors on $\dot{M}$

The high spin model does result in slightly lower $i$ and $D$ and slightly higher $M_{BH}$ compared to the low spin model. The lower $i$ value from the high spin model is in better agreement with the more recent modelling of jets from \mx\ (Carotenuto et al. in prep.).

Considering obs1 and obs2, the low and high spin models give similar results: they give similar quality fits at the posterior mode (see above) and similar marginal distributions on most parameters. For a given model, the posterior for $\dot{M}$ differs between observations in line with changes expected for the different X-ray fluxes of the observations (e.g. \swift/XRT count rates shown in Fig. \ref{fig:DATES_vs_FLUX}).

\begin{figure*}
	\includegraphics[width=16cm]{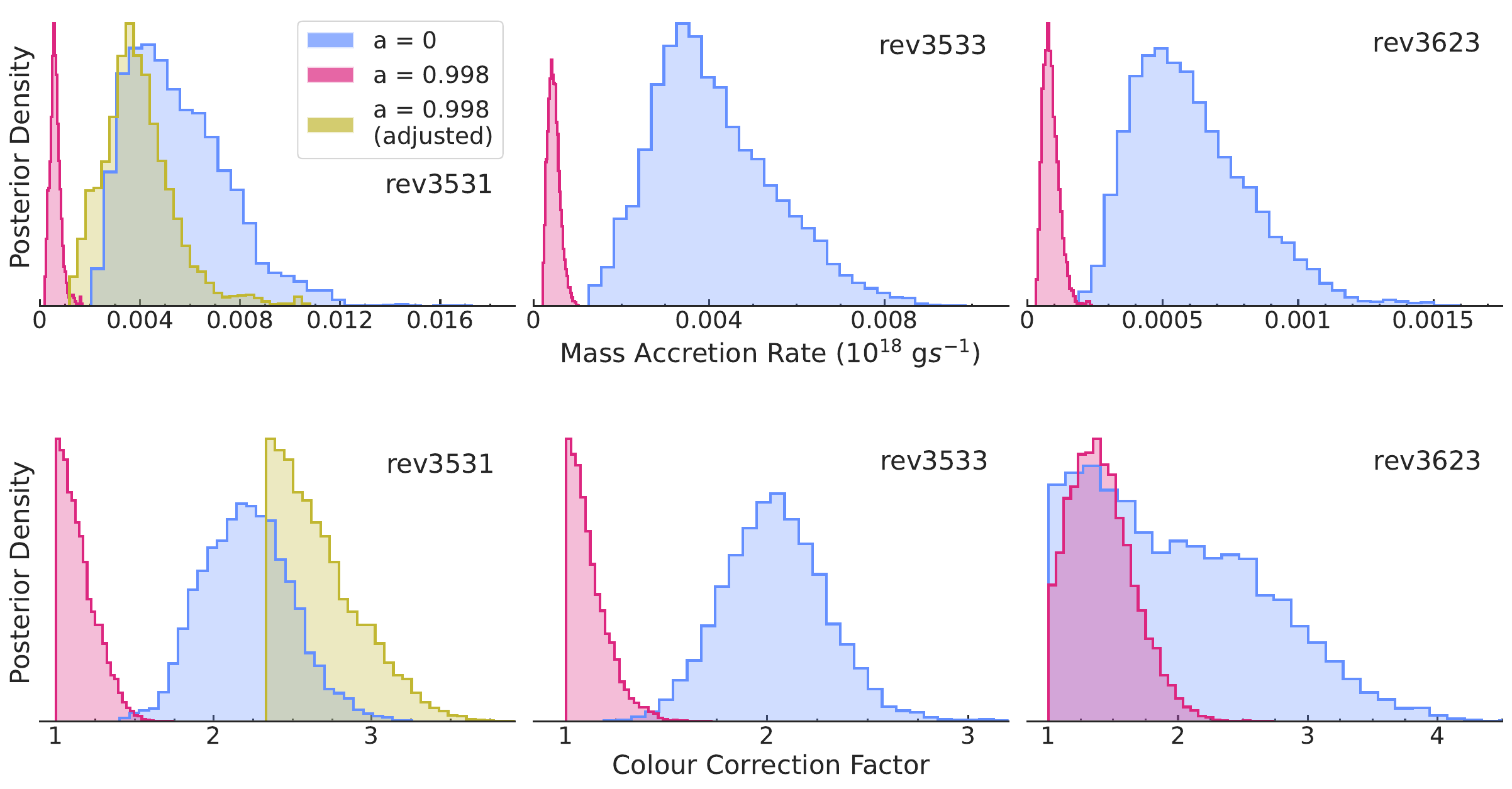}
    \caption{Posterior distributions for the mass accretion rate (top) and the colour correction factor (bottom). The low and high spin model posteriors are shown in blue and pink respectively. 
    For obs1, we show in yellow the posteriors from the high spin model scaled by factors  $\sim 6.4$ (for $\dot{M}$) and $\sim 2.3$ (for $f_{col}$). The posteriors are in much closer agreement once adjusted for this change expected from the difference in radiative efficiency between the two models. We found the same for the other two observations (not shown here).}
    \label{fig:M_dot_fcol}
\end{figure*}

There are two parameters that show markedly different posteriors in low and high spin models: $\dot{M}$ and $f_{col}$. See Fig. \ref{fig:M_dot_fcol} for the MCMC posterior distributions (the same discrepancy is also seen for the posterior distributions from NS). This is expected due to changes in radiative efficiency ($\eta$) between the two models.
The distance prior and the flux of the data are the same for each model, but $\eta$ is spin-dependent: for the $a=0$ model $\eta \approx 5.7$\%, whilst for $a=0.998$, $\eta \approx 32$\% \citep{Thorne_1974}. Therefore, to match the same flux at the same distance, the $a=0.998$ model must have a lower accretion rate by a factor of $\sim 32/5.7 \sim 6$. This also changes the disc temperature profile, but this can be compensated for by a change in $f_{col}$. Adjusting for this expected difference (see Fig. \ref{fig:M_dot_fcol}), the two posteriors for the mass accretion rate come into closer alignment with one another\footnote{In order to demonstrate this, we simulated a spectrum with an extremely long exposure from \textsc{KERRBB}, for similar system parameters to those of our model. Fitting $a=0$ data with an $a=0.998$ model, resulted in a reduction in the colour correction factor and mass accretion rate of $\sim 2.3$ and $\sim 6.4$ respectively. Adjusting our $a=0.998$ model posterior samples for $\dot{M}$ and $f_{col}$ by these factors, we find a good agreement between the posteriors for $a=0$ and $a=0.998$ model in all but the colour correction factor of obs3, where the there is already a large overlap.} We conclude that -- apart from the expected changes in $\dot{M}$ and $f_{col}$ -- the low and high spin models can hardly be distinguished on the basis of their parameters. Worth noting is that $f_{col}$ is limited to $\ge 1$ and for the high spin models the posterior samples are close to this limit, which truncates the distribution.

\subsection{Reflection iron abundance}

In all our modelling, we find very high iron abundances in the disc, with $A_{Fe} \sim 10$. Super-solar iron abundance is often reported from X-ray spectral modelling of both XRBs and active galactic nuclei \citep{Parker_2016, Ludlam_2020, Dong_2020, Garcia_2018}. Previous modelling of \mx\ with \textsc{relxill} also yielded high iron abundances. \cite{Chakraborty_2020} found iron abundances in the range $4.4 - 5.0$ times solar, from \nustar\ data.

As discussed in \cite{Garcia_2018_P}, such high values may be an artifact of inaccuracies in modelling the reflection spectrum. In particular, real XRB discs are likely to be more dense than the plasma modelled by \textsc{relxillCp} (model limit of $n_e$ = $10^{20}$ cm$^{-3}$). Further, slight inaccuracies in the atomic physics and high density and temperature may result in biased abundances from model fitting.

\subsection{Reflection emissivity}

For the two emissivity indices, we find good agreement between the posteriors in all three observations for both high and low spin models. For $I_1$ (inner emissivity, between $R_{br}$ and $R_{ISCO}$), the posterior is similar to the prior, suggesting the data do not carry much information about this parameter; in each case the posterior peak is only slightly below the prior peak (at $I_1 = 3$). 
The situation is different for $I_2$ (emissivity between $R_{out}$ and $R_{br}$). The prior peaks at $I_2=3$ but the posterior peak lies in the range $1-2$ (i.e. a flatter emissivity law in the outer disc). An emissivity with $I_2 <2 $ gives increasing reflection luminosity at larger radii, i.e. reflection is dominated by emission from far out in the disc. 
This is contrary to the expectation of simple models with a centrally-illuminated flat disc, but could be consistent with a flared or warped disc. However, the reflection features in these data are relatively weak; the Fe-K$\alpha$ equivalent width is $\sim 65-90$ eV when fitted with a simple Gaussian line and the reflection peak $>10$ keV is hard to discern in the data. Together with the known limitations of the {\sc relxillCp} model we advise caution in interpreting the posterior results for the reflection parameters. 

\subsection{Obs3 data} 
\label{rev3623}

The data for obs3 pose additional challenges to interpretation. Obs3 is the faintest of the three sets of observations, at just below $\sim 2 \times 10^{-4}$ $L_{Edd}$ and 
the fit statistics at the posterior mode are slightly worse that for the other two observations. Furthermore, there are significant disagreements between the EPIC pn and \nustar\ data where they overlap $\sim 5-10$ keV (see Fig.~\ref{fig:RESIDUALS_rev3623}). We also find substantial differences when comparing posterior samples from NS to those from MCMC, see \ref{tab:Param_values_NS}. Stark differences exist for many of the free parameters, including $N_H$, $\Gamma_S$, $log(\xi)$ and $A_{Fe}$. Given these issues in fitting the data and obtaining reliable posterior samples, we must treat inferences from obs3 with caution.

\subsection{Bayes Factor}

\begin{table*}
	\centering
	\caption{A comparison of BF for all three revolutions. The `SDDR' values use the Savage-Dickey Density Ratio, and the `evidence ratio' values are the differences in $\log(Z)$ (marginal likelihood from NS). In all cases, the BF is a comparison between a restricted model with $a$ fixed (at either $0$ or $0.998$) and a more general $a=$ free model, and values are reported as $\log(BF)$ (based $10$) for ease of interpretation. We computed each SDDR using two different estimates of the posterior density, one from MCMC and one from NS. Errors on $\log(Z)$ are omitted but small, of order $10^{-2} - 10^{-1}$.
 }
	\label{tab:Bayes_Factor}
	\begin{tabular}{r|cc|cc|cc}
		\hline 
		\hline
         & \multicolumn{2}{c}{obs1} & \multicolumn{2}{c}{obs2} & \multicolumn{2}{c}{obs3}\\
        & a = 0 & a = 0.998 & a = 0 & a = 0.998 & a = 0 & a = 0.998 \\
        \hline
        MCMC BF (SDDR)           & 0.09   & -1.07 & -0.02  & -3.02  & -0.05  & -1.41\\
        \hline
        $\log(Z)$                & -48.24  & -49.79 & -142.75 & -144.80 & -151.86 & -151.24\\
        NS BF (Evidence Ratio)   & 0.04    & -1.51  & -0.05   & -2.11    & 1.72    & 2.34 \\
        NS BF (SDDR)             & 0.04   & -3.92 & 0.10   & -3.12   & 0.26  & -7.02\\
		\hline
		\hline
	\end{tabular}
\end{table*}

\begin{figure}
	\includegraphics[width=\columnwidth]{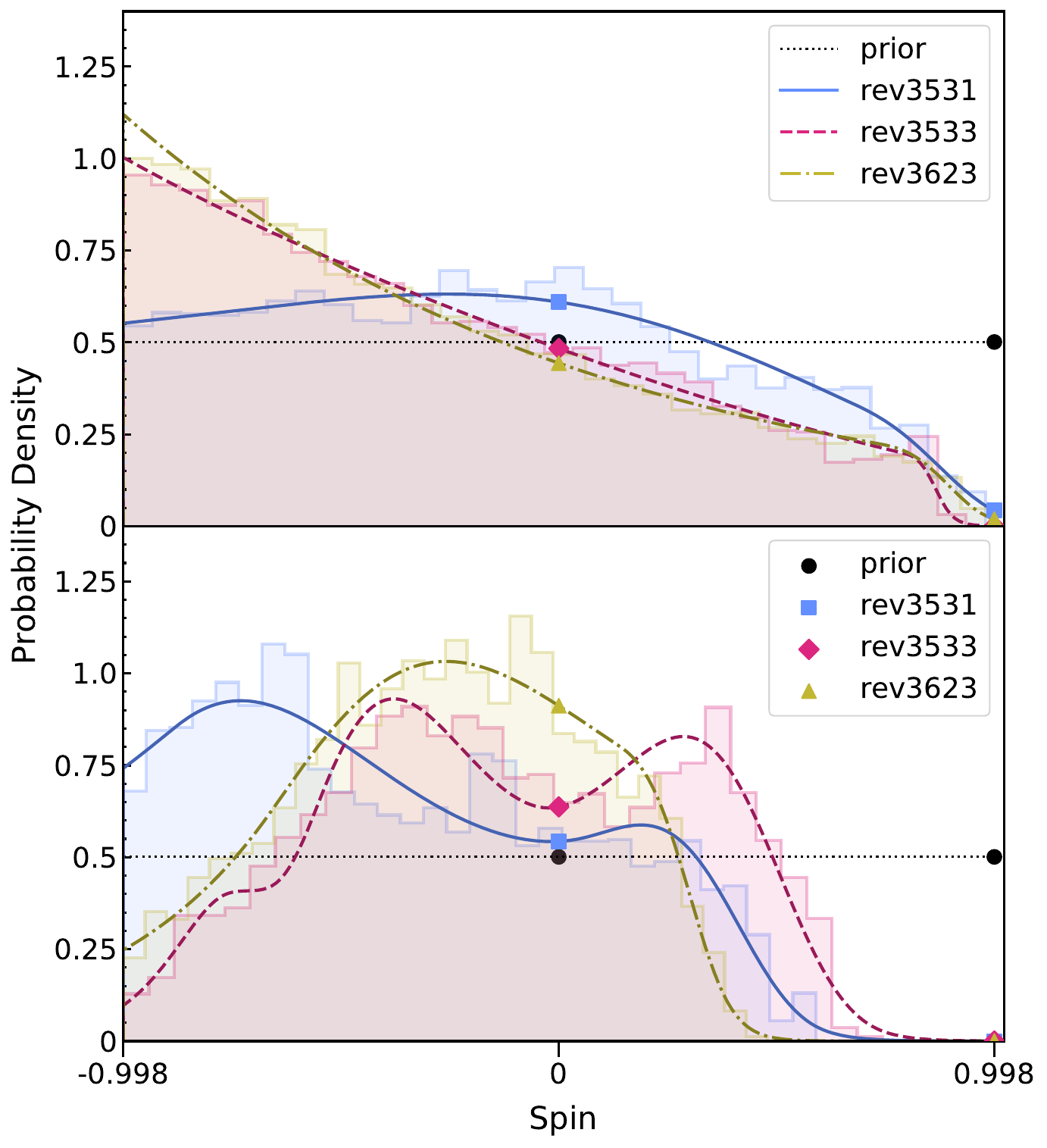}
    \caption{Distributions of the posterior probability densities of spin from MCMC (\textit{top}) and NS (\textit{bottom}). Obs1, obs2, obs3 and the uniform prior ($\approx 0.5$), are shown by the solid blue, dashed pink, dashed-dotted yellow and black dotted lines respectively. Lines were obtained using a logspline density estimation \citep{logspline_1997} for a maximum of six knots. Histograms for the data are shown in the same colour as each line. Also in the same colour scheme are circles, squares, diamonds and triangles representing the height of the probability density at $a = 0$ and $a = 0.998$ for the prior, obs1, obs2 and obs3 respectively. The range of values for both plots are the same and between 0 and 1.4.}
    \label{fig:SPIN_MCMC_NS}
\end{figure}

The posterior density estimates for parameter $a$ are illustrated in  Fig.~\ref{fig:SPIN_MCMC_NS} based on marginal posterior samples of $a$ from MCMC and also from NS. These show a relatively high posterior density at low to moderated values of $a$, decreasing as $a$ approaches $0.998$. 

Here, we apply methods of Bayesian model comparison to assess the models (see Appendix \ref{Analysis_Methods} for more details of the methods). 
We take the special cases of a zero-spin ($a=0$) and a maximally spinning ($a=0.998$) and in each case compare it to the more general free-spin model. The special cases are nested models within the more general model. The results of two different methods are summarised in Table~\ref{tab:Bayes_Factor}. 

The BFs computed using the MCMC posterior estimate shown in Fig.~\ref{fig:SPIN_MCMC_NS} (top panel) -- using the SDDR method -- give a preference for low spin from all observations, although much more strongly in the case of obs2 than obs1 and obs3. 
We found $BF=1.217$ for $a=0$ from obs1, i.e. no preference between the $a=0$ (fixed) model and $a=$ free model, while $BF=0.086$ for $a=0.998$ modestly favours the $a=$ free model over the $a=0.998$ model by odds of $\sim 12:1$.

The BFs computed by NS -- the ratio of marginal likelihood estimates estimated by the sampler -- also favour the low spin case for obs1 and obs2. In contrast, in the case of obs3, we find a model preference of $\sim$ $4:1$ in favour of high spin.
If taken literally, this result is below significance \citep{Practical_Bayesian_Inference_Book}, and we caution against any interpretation of the BF for this observation. Obs3 is the observation with largest disagreement between the MCMC and NS methods (see Section~\ref{rev3623}), and so we suspect that the one or both methods have failed to converge as expected. Results from using NS with an even larger number of steps supported this. The apparent preference for the high spin model runs against the natural inference from the NS posterior samples shown in Fig.~\ref{fig:SPIN_MCMC_NS} (bottom panel). 

Repeating the SDDR method with the NS posteriors, we find a very strong preference in favour of low spin. 
There are so few NS samples at high spin that it is difficult to estimate the BF with this method -- the highest $a$ values in the posterior samples are $0.59$, $0.75$ and $0.48$ for obs1, obs2 and obs3, respectively. This itself is an indication that the high spin models are not favoured, but makes an estimate of the BF difficult. Assuming there is a single count in a bin above these upper limits, we still find BFs $10^3$ - $10^4$ times smaller for $a=0.998$ than for $a = 0$.

\section{Discussion}
\label{disucssion}

We have demonstrated the use of Bayesian modelling to study X-ray spectra from the black hole X-ray binary \mx. In this section we review this approach, contrast with alternative approaches, and discuss the implications for understanding \mx. We divide the discussion into two main parts: first, a discussion of the model, its assumption and use in practice, then a discussion of the interpretation of the results of fitting the model to the \mx\ data.

\subsection{The model}

\subsubsection{Shape of the model}
\label{assumptions}

Our underlying model, which we call the {\it vanilla} model, is of a conventional accretion disc (flat, geometrically thin and optically thick), in the plane perpendicular to the axis of the jet, and that extends inwards to approximately $R_{ISCO}$ with a zero torque inner boundary condition. The disc emits a thermal spectrum, and hard X-rays are produced by inverse-Compton scattering of seed disc photons that in turn reflect off the ionised surface of the disc. The exact choice of X-ray spectral model components that we combine to approximate the X-ray spectrum from such a system is discussed in Section \ref{Vanilla_model}. 

\subsubsection{Priors}
\label{disco_priors}

A key step in any Bayesian workload is defining reasonable priors for the model parameters. Our model has many ($22 - 24$) free parameters, each has been assigned a prior density; we crudely classify each as either `informative' or `non-informative'.

The parameters assigned informative priors are those that are well understood based on previous studies. A few of the model's parameters represent the fundamental parameters of the binary system: $M_{BH}$, $i$, $D$. \mx\ is now a well-studied binary system, and so for these parameters we define informative priors based on previous multi-wavelength studies. See Section \ref{sect:binary}. The black hole mass, $M_{BH}$, is not directly estimated from multiwavelength studies, but a prior distribution can be formed based on the available measurements. An accurate model of X-ray absorption by intervening ISM is an important element of such a model, and here we use high-resolution RGS data to constrain the column densities of eight neutral and weakly ionised species. The results of the RGS modelling are then used to form priors for the column densities using in the subsequent EPIC and \nustar\ modelling. See Section \ref{RGS_section}.

The remaining parameters are assigned non-informative priors. These are the parameters for which we do not have detailed information, and so we define broad priors spanning ranges that may be limited by theory, by the restrictions implicit in the vanilla model prescription, by previous observations the ranges seen in other X-ray binaries, or by the computational limits of the model codes being used. These include: $\Gamma_S$, $\Gamma_R$, $f_{col}$, $f_{scat}$, $I_1$, $I_2$. See Section \ref{priors}.

One aspect of our model that is different to many (but not all) previous models is that we allow the photon index of the primary hard X-rays ($\Gamma_S$) and of the reflection spectrum ($\Gamma_R$) to be independent parameters. Specifically, we assign $\Gamma_R$ the same broad prior as used for $\Gamma_S$ but the two parameters are not tied together in the modelling process. We thereby relax the common assumption that the hard X-ray spectrum incident on the inner disc is the same as the spectrum seen by us as distant observers. This also allows the reflection model some degree of flexibility to absorb biases caused by inaccuracies or incompleteness in the physics included in the reflection emission model. There are some parameters that may plausible span a very large range, or for which we have very little theoretical understanding and there is no consensus in the literature. To these, we assign very broad priors: e.g. uniform priors over the full range allowed by the available model for $\ln(\dot{M})$, $A_{Fe}$, $\log({\xi})$, $\ln(norm_R)$. 
(The use of different log bases here is due to the way the parameters are specified {\tt XSPEC} and has no effect on the shape of the posterior.)
There are also two parameter used to cross-calibrate the fluxes between the three spectra. We apply simple uniform priors for the flux scaling parameters used between \xmm\ EPIC and \nustar\ FPM-A and FPM-B spectra. 

\subsubsection{Review of the assumptions}
\label{disco_assumptions}

The vanilla model is built on the assumption that $R_{in} \approx R_{ISCO}$, i.e. that there is no (or very limited) truncation even in the hard state at modest luminosity. One motivation for this is to test whether a model that takes this assumption seriously can fit the data and give values for the other parameters that are consistent with what is known about the system and with the best theoretical arguments about the origin of the emission. Our inferences about the black hole spin (see below) are conditional on our assumptions, and particularly the assumption of little or no truncation of the disc outside of $R_{ISCO}$. 

As discussed in Section \ref{Introduction}, one popular model for state evolution in X-ray binaries has the thin disc truncating outside of $R_{ISCO}$ as the accretion rate drops toward quiescence. \cite{Xu_2020} modelled the reflection spectrum of obs3, finding $R_{in} \gtrsim 30$ $r_g$, concluding the disc is truncated in agreement with the popular model. Their analysis involved very different assumptions to ours. They do not make use of information about the jet inclination to inform the disc inclination parameter, and in several of their models their best fitting inclination, based only on X-ray spectral modelling, is inconsistent with the inclination of the jet that we make use of. Their emissivity model is a power law of index fixed at $3$, and other parameters have implicitly flat priors (no explicit prior densities defined).

Although disc truncation is not included in our model, certain combinations of parameters could approximate some of the effects of a modestly truncated disc, e.g. $R_{in} \sim 20 r_g$. A model with $a \approx -0.998$ would give a similar thermal disc spectrum to a truncated disc around a black hole with higher $a$. A lower $f_{col}$ could further mimic the effect of truncation by shifting the disc spectrum to lower energies than expected for $R_{in} \approx R_{ISCO}$. A very flat inner reflection emissivity index $I_1$ could mimic the effect of a truncated disc by removing most of the reflected flux from $<R_{br}$. The strongest indicator of a failure of our assuming of $R_{in} \approx R_{ISCO}$ would be a very poor fit to the data. None of the above possible signals of truncation were seen in our analysis. It difficult to reconcile our results with the \cite{Xu_2020} claim of $R_{in} \gtrsim 300 r_g$, for inclinations comparable with the prior information from the jet, but our results are more consistent with the radii found in \cite{Buisson_2019} ($R_{in} \sim 5 r_g$). However, these observations were taken during the main outburst, with higher luminosities ($L \sim 0.01 - 0.1 L_{Edd}$), making comparisons to $R_{in}$ more dubious.

Our assumption that the plane of the inner accretion disc is perpendicular to the axis of the jet is in line with expectations of the vast majority of theoretical models. The jet axis is usually thought to be defined either by the plane of the inner disc or by the spin axis of the black hole (if high spin). In the latter case, one expects the Bardeen-Petterson effect to align the inner disc perpendicular to the black hole spin, and hence also the jet. 
As mentioned in Section \ref{sys_params_inclination}, \cite{Poutanen_2022} suggested that the binary orbit and jet were misaligned. Here we find the posterior for the disc inclination is broadly consistent with the simple model where the jet axis is perpendicular to the disc plane, in line with our assumption.

Our model assumes the disc is continuous, planar (no warps or bulges), at least in the inner regions where the thermal emission dominates. Our broad priors on the reflection emissivity of the `outer' disc (which we define as $R > 18r_g$) allow for discs that subtend more or less than the sold angle expected for a flat disc, and thereby allow for some degree of warping or flaring in the outer regions. 

Our vanilla model assumes that the primary hard X-ray photons come from inverse Compton up-scattering in a corona. The detailed geometry of the corona is left undefined\footnote{The reflection spectrum normalisation is a free parameter (not tied to the flux of the primary hard power law), as are the two power law indices for the reflection emissivity law. }, and the primary implication of this assumption is that the photon number is set by the thermal disc emission and conserved as photons are boosted to hard X-ray energies to form the hard X-ray power law.  
The scattering fraction (from soft, thermal X-rays into hard, non-thermal X-rays) can vary over the interval $[0,1]$ and we assign a uniform prior to this parameter. If there are other sources of hard X-rays (e.g. shocks) then this assumption will not be valid, and the effective $f_{scat}$ could exceed $1$.

Another assumption we make is that there is zero torque ($\eta$ = 0) at the inner edge of the accretion disc. This modifies the thermal spectrum of the inner disc, but the possible effects of changing this parameter are much smaller than those from changing $\dot{M}$ or $f_{col}$ - see e.g. Figs. 6, 7 and 8 of \cite{Li_2005}. 

Lastly, we fixed the emissivity break radius. In reality, the reflection emissivity may not be a single power law but could be approximated by piecewise power law functions, here we fix a break at $R_{br} = 18$ $r_g$ and allow for a different power law index at smaller ($I_1$) and larger ($I_2$) radii. We assume the reflector extends from $R_{in}=R_{ISCO}$ (set by $M_{BH}$ and $a$) and out to $R_{out} = 1000$ $r_g$ (the limit of the model).
In order to manage the computational issues (see Section \ref{Practicalities}), we did not allow $R_{br}$ to be a free parameter. The exact placement of $R_{br}$ will have a stronger effect on the resulting spectrum if there is a large difference between the indices $I_1$ and $I_2$ and the radii near $R_{br}$ dominate the total reflected emission. These appear not to be the case for the values we obtained by fitting to \mx. 

The reflection model assumes a high-energy cut-off in the incident power law spectrum. We fixed $kT_e = 400$ keV. Based on previous work on \mx, there is little evidence for a cut-off within the energy range likely modify the reflection spectrum over our observed bandpass, and so we fixed this parameter (to limit the number of free parameters) at a value well outside of our observed bandpass (approx $0.8-70$ keV). 

\subsubsection{Relaxed Assumptions}  
\label{relaxed_assumptions}

A common assumption in modelling the spectrum of X-ray binaries is that the disc atmosphere modifies the effective temperature of the thermal emission by a factor $f_{col} \approx 1.7$ \citep{Shimura_1995}. We allow for a much wider range of values for this parameter using a broad prior spanning the interval $[1,10]$ but peaking at $\approx 1.7$. \citet{Salvesen_2013} discussed some ways in which higher values of $f_{col}$ might be realised and important to consider for X-ray binaries. 

Although our vanilla model is designed to approximate the emission expected for a thin, flat disc, we implicitly allow for some deviations from this by allow for a range of inner and outer reflection emissivity laws and for differences between the observed power law and that incident on the reflector. 

In addition to relaxing the above assumptions, we adopt fully Bayesian methods for dealing with both ISM column densities and system parameters. We use broad column density priors reflecting the depths of absorption features inferred from \xmm\ RGS data and system parameters priors created from the best optical and radio data. This removed the need to assume a fixed value from the literature for parameters such as the black hole mass or $N_H$ (e.g. from $21$ cm radio maps). We contrast this with two alternative approaches that are commonly employed in this field. 

The first approach is to make hard assumptions about these parameters -- e.g. fixing $D$, $M_{BH}$, $N_H$ and implicitly assuming ISM abundances (e.g. default values in models). In Bayesian terms, this is using a delta function prior on parameters. The results of such analyses are conditional on the appropriateness of the very strict assumptions. As the X-ray spectra are highly sensitive to some of the parameters (e.g. $N_H$ or $D$), we should trust inferences based on such modelling only as much as we trust the validity of the strict assumptions used. 

The second approach that is often employed is to leave all or most model parameters free with no explicit specification of a prior. In Bayesian terms, this is implicitly applying a uniform prior on all free parameters. This avoids the issue of strict assumptions, but also fails to make use of the sometimes very good information we have about the system (e.g. from non X-ray studies). The result can be that many -- sometimes quite different -- combinations of parameters give comparably good fits and one cannot distinguish between them (because pairs of parameters are degenerate in the modelling). Making use of strong prior information can help to reduce these issues without imposing unreasonably strict assumptions about parameters. 

As an experiment, we removed all our priors -- implicitly setting them to uniform over a wide range -- and fitted the model to the EPIC pn and \nustar\ data in order to examine the extent to which our priors and Bayesian analysis effected the posterior distributions. The medians of most column densities and some system parameters varied significantly from their Bayesian counterparts. The most effected of these was the mass of the black hole; in several combinations of data (three observations) and model ($a=0$, $0.998$, free) we found $M_{BH}> 10 M_{\odot}$. There was also a large differences in the inclination, with best fitting values from $i \approx 45^{\circ}$ to  $75^{\circ}$, and distances differing between observations over the range $D=1.3-2.7$ kpc. These have effects on the other parameters, e.g. the lower $D$ causes a reduction $\dot{M}$ and $f_{scat}$ and an increase in the colour correction factor to compensate. 

\subsubsection{Practicalities} 
\label{Practicalities}

Where we do make strong assumptions about model parameters, e.g. some are fixed or restricted to a limited range, these are largely made for practical reasons. 
In some cases, the code to generate the model spectrum has limits on the allowed values of parameters. Other parameters were expected to have little effect on the inferences -- e.g. because the spectra model over the bandpass of our observations is insensitive to the expected range of the parameter -- and these were fixed at plausible values in order to limit the number of free parameters (see Section \ref{disco_assumptions}). Each additional parameter added to the time needed to reach convergence for both MCMC and NS methods. 

\subsubsection{Self Consistency} 
\label{self-con}

One intention of our approach to model design was to impose as much self-consistency between model components as possible given the available model components. The motivation is to leverage physical consistency between some parameters to improve the reliability of inferences about the model. 

One common approach to modelling X-ray binary spectra is to add emission models for the thermal (disc) and non-thermal (power law) components, with independent normalisations, effectively allowing the luminosity of each component to vary independently of each other. If the power law photons originate in the disc and are inverse-Compton scattered into hard X-rays, the disc emission and $f_{scat}$ determines the total number of hard X-ray photons, conserved by the scattering process. The luminosity of the disc (which in turn is set by the accretion rate) is thereby coupled to the luminosity of the power law. The observed flux of the X-ray power law, is thereby related to the disc luminosity through the $f_{scat}$ parameter. 
In the extreme case, even if the thermal disc emission is too cold or faint to observe directly in the X-ray spectrum, the hard X-ray luminosity (together with strong prior information on $D$, $M_{BH}$ and the non-detection of thermal emission) constrains the accretion rate; this is possible because we model the luminosities of the thermal and non-thermal components as physically related.

A further issue with the simple addition of a hard X-ray power law is 'flux stealing', as discussed in Section \ref{Vanilla_model_components}. The extrapolation of the scattered spectrum to energies below those of typical seed photons will overestimate the non-thermal flux at low energies, and so underestimate the thermal flux, and this will affect the inferred accretion rate and other parameters. 

Our composite model is not fully consistent on this front. We used the convolution model \textsc{SIMPL} for the hard X-ray power law, which drops off at lower energies and creates the inverse-Compton spectrum by removing photons from \textsc{KERRBB}, which represents the thermal, disc emission. There is therefore consistency between these two model components. The reflection spectrum -- modelled using \textsc{relxillCp} -- is a simple addition to the total flux, there is no consistent treatment of thermal reprocessing of hard x-rays on the disc emission. 

The geometry of the reflector is tied to that of the thermal disc to the greatest degree possible, to maintain geometrical consistency. The inclination parameters of each component are tied together, as are the black hole spin parameters. Both models include emission down to $R_{in}=R_{ISCO}$. As discussed above, we did not tie together the photon indices of the directly observed power law ($\Gamma_S$) and power law of the spectrum assumed incident on the reflector ($\Gamma_R$). We effectively treat $\Gamma_R$ (or the difference, $\Delta \Gamma = \Gamma_S - \Gamma_R$) as a nuisance parameter (see Section~\ref{Vanilla_model_components}).

\subsubsection{Comparing MCMC and NS}

Extracting inferences from this model is computationally challenging. There are more than twenty parameters in the main model, and sampling from the posterior distribution proved to be time consuming. We used an MCMC method and NS, and compared the outputs of each. Using data from obs1 or obs2, we found an agreement for almost all parameters with an overlap in their errors and largely similar distributions, with only a few exceptions. There was minor disagreement in the location of the peak of the scattering fraction ($\sim 10 - 15\%$). We currently have no explanation for why only this parameter varied between sampling methods, however the differences are small and do not affect the results. Overall, the posteriors from both methods agreed well, despite using two different fit statistics ($\chi^2$ for MCMC; $W$-stat for NS). This consistency suggests that each sampling method has converged, at least for modelling of obs1 and obs2. The methods produced posterior estimates that were quite different in the case of obs3, suggesting that at least one method could not converge even with the (very large) number of iterations used; exactly why obs3 proved more challenging remains unclear. 

MCMC typically took longer to reach convergence (by a factor of $\sim 1.1 - 1.8$) than NS. This assumes MCMC reached convergence after 20 batches (see Appendix \ref{Criteria_for_convergence}), and depends on the step size used for drawing new `live' points for NS (see Appendix \ref{NS_Appendix}). This comparison of computational speed is specific to our choice of model, priors, and also to the resolution of the data being fitted. Changes to any of the above may result in significantly different convergence times for either method.

\subsection{Inferences from modelling \mx}

We begin this subsection with a few remarks about the model as applied to \mx, and then discuss in more detail our results for two of the more interesting parameters: $f_{col}$, $a$.

Overall, the vanilla model works well in the sense that it provides a good fit to the data in all three observations, over the $0.8-50$ keV range. See e.g. Figs.~\ref{fig:RESIDUALS_rev3531} $-$ 
 \ref{fig:RESIDUALS_rev3623}. The posterior for some key system parameters ($M_{BH}$, $D$, $i$) are consistent with the strong prior information. This means the modelling process is not able to recover much information about these from the data, but by using this information we are able to ensure the model agrees with established results about the binary system. 

The posterior distributions for the reflection parameters reveal a weak reflection component, and a high iron abundance. This is commonly found in spectral fitting of X-ray binaries (and also AGN) but is not well understood. The emissivity function for the reflector is superficially reasonable, modelled with a broken power law that is steeper in the inner regions. We note that the emissivity index of the outer reflector ($r \approx 18-1000$ $r_g$) is below $2$, meaning the reflection is dominated by outer regions. This may be an artefact caused by some deficiency in the reflection model but if real could be explained by a flaring of the outer disc. 

The scattering fraction needed to explain the hard X-ray power law is $f_{scat} \approx 0.7$. We would expect this to be high (slightly below $1$) for in the hard spectrum / non-thermal state where the coronal/jet emission dominates. 
Our estimates of the broad-band X-ray luminosity are $10^{-4} - 2 \times 10^{-3}$ of the Eddington limit. We would expect values $\ll L_{Edd}$ given the source is in a hard state and the X-ray luminosity is more than a factor $10^2$ smaller than the outburst peak. 
The mass accretion rate required to power the thermal disc emission can be inferred reasonably well despite the fact that the thermal disc emission is very weak because of the assumption that hard X-rays are up-scattered soft photons originating in the disc (but see the caveat below). Essentially, the fact that photon number is conserved in this process -- combined with reasonable knowledge of $M_{BH}$ and the assumption of $R_{in} \sim R_{ISCO}$ -- allows the model to constrain the disc luminosity and hence mass accretion rate needed to produce the seed photons, even when the thermal disc emission is weak. Of course, these assumptions may not apply to this system, but this work demonstrates how an understanding of the physical connection between radiation processes in these systems can be leveraged to make inferences about physical parameters that are otherwise hard to constrain. As the total X-ray luminosity is dominated by the non-thermal power law, the total mass accretion rate must be much larger than that inferred from the {\sc KERRBB} model component alone. 
Our estimate for the total X-ray luminosity is $\sim 7$ larger than the luminosity inferred from the model's accretion rate.

\subsubsection{Colour Correction Factor}

The `hardening' or 'colour correction' parameter for the disc emission was $\approx 2-3$ for most of our models. This is higher than some theoretical work predicts, but well within the ranges discussed by e.g. \citet{Salvesen_2013}. The high spin model (with $a=0.998$ fixed) gave the smallest $f_{col}$ values, in some cases reaching the minimum allowed value of $1$, leading to a posterior that is truncated at the lower end. If taken literally, this could be an argument against the high spin model, which seem to require an unreasonably low value for this parameter. Of course, this inference is conditional on the optically thick disc emission extending to $R_{in} \approx R_{ISCO}$. 

\subsubsection{Spin}

The overall shape of the posterior distributions on the black hole spin parameter, $a$, from MCMC (see Fig. \ref{fig:SPIN_MCMC_NS}) is quite flat over most of the allowed range of $a$, and suggests that the model is largely insensitive to a change in spin. For each of the three observations, the marginal posterior for $a$ has a higher density for lower spin (e.g. $a < 0.5$), although the details vary between the observations. The posterior drops as $a$ approaches is maximum ($0.998$), suggesting that the model has difficulty with higher spins. The shape of the marginal posterior for $a$ is different between NS and MCMC analyses (see Fig. \ref{fig:SPIN_MCMC_NS}), but these general conclusions remain consistent, the difference is in the extent of the drop in posterior density at highest spins. 

The most straightforward conclusion is that a low or moderator spin is preferred over very high spin. Again, this conclusion is conditional of the assumptions inherent in the vanilla model. The most important here is that  $R_{in} \approx R_{ISCO}$. If the disc was truncated ($R_{in} \gg R_{ISCO}$), then a higher spin could be accommodated as the effect of $a$ on the disc emission at large radii is much smaller. 

Our results agree with the claims of low spin in the literature \citep{Guan_2021, Zhao_2021, Buisson_2019}. 
That said, some analyses of \mx\ do claim high spin, e.g. \cite{Draghis_2023}. 
\cite{Bhargava_2021} used X-ray timing data and fit a model for relativistic precession to claim $a\approx 0.8$. It is difficult to reconcile these results with ours given the current data.

\subsubsection{Remarks on \mx}
\mx\ is one of only 11 confirmed black hole LMXBs, with a relatively well constrained $M_{BH}$ and $i$ and one of the most accurate parallax measurements ($D$) of any BH LMXB. It is additionally one of the few black hole XRBs (both confirmed and candidate) with good RGS and \nustar\ data in a hard state -- the state in which disagreements about whether truncation occurs are important. Earlier (much brighter) parts of the outburst of \mx\ did not yield usable \xmm\ data due to pile-up. This prevented us from being able to test the {\it vanilla} model on soft state observations using \xmm. We plan to extend our modelling to others LMXBs and to soft state observations to better test the {\it vanilla} model.

\subsection{Opportunities for further work}

There are several ways in which the model and method described above could be improved. First, there are physical processes that are not included in the spectral model. Thermal reprocessing is neglected; the disc surface is irradiated by hard X-rays, some fraction of this energy goes into the reflection spectrum, but the rest is absorbed and raises the temperature of the disc surface above that expected solely from internal heating due to the accretion process.  Our estimates of $\dot{M}$ may be overestimates if this is an energetically important effect for \mx\ during these observations. Another limitation is that the {\sc KERRBB} model is based on the standard theory of a radiatively efficient disc and this may not extend down to the low accretion rates implied by our modelling. We do not account for the heating of the hot electrons (in the corona). Some accretion power must go into energising the corona, but our model assumes all the accretion power is radiated by the disc. This could lead to our model underestimating the true $\dot{M}$. These effects pull in opposite directions and it is not clear which dominates without a fully self-consistent theory for the disc-corona coupling that can extend over a wide range in luminosity. 

A more advanced model could explicitly include truncation of the disc -- allowing $R_{in}$ to vary would affect both the thermal disc emission \citep[see e.g.][]{Hagen_2023_RELAGN, Dovciak_2004_KYN} and the reflection spectrum. This would require additional parameters, and so increase the computation challenges of generating posterior inferences from the model. But, at least in principle, this would allow for examining both black hole spin and truncation together. 

We did not include dust (molecular components) in the model for ISM absorption; this could have an effect on the detailed shape of the spectrum at low energies and, if not accounted for, could slightly bias some of the other parameters. A more comprehensive model could account for all these effects, but at the cost of additional parameters and hence more challenging computations for sampling the posterior.  

We used independent priors; the joint prior for all parameters is simply the product of the prior assigned to each parameter. Some parameters could have a joint prior. e.g. the pair $M$ and $i$. We used posterior distributions for the mass function ($f$) and mass ratio ($q$) to generate a joint distribution for $M$ and $i$. These parameters are correlated, but we did not make use of this information; instead, we used the product of the marginal distributions for $M$ and for $i$ which is more diffuse over the $M-i$ plane. Using the joint prior and accounting for the dependence of these parameters would have been more informative and helped to further restrict the range of the posterior. 

We could also extend the fitting over a wider energy range, in particular, making use of the EPIC pn data down to e.g. $0.5$ keV. We restricted the range used in this study in order to limit the impact of small systematic errors in the calibration of the instrument response or in the ISM absorption model, which can have a much stronger effect on the spectrum at lower energies.  

\section{Conclusion}

We analysed three \xmm\ and two simultaneous \nustar\ observations from the LMXB \mx, applying a fully Bayesian approach to the modelling problem with the {\it vanilla} model. We tried variations of the model with low and high spin black hole spin, as well as spin as a free parameter. The design goals of the modelling approach were:

\begin{itemize}
\item{Where possible, apply self-consistency between the model components.}
\item{Informative priors on column densities for ISM absorption features modelled based on high-resolution RGS data.}
\item{Informative priors on the binary system parameters ($M_{BH}$, $i$, $D$) from the best optical and radio measurements in the literature.}
\item{Relaxed assumptions on other parameters where the physics is not well understood, such as the accretion disc colour correction factor.}
\end{itemize}

This is the first, to our knowledge, application of a fully Bayesian workflow to the problem of X-ray binary spectra. Past work, such as \citet{Connors_2019} includes some aspects of the Bayesian workflow. From our analysis we find the model works well in the sense that the quality of fit to the data (e.g. $\chi^2$) is reasonable and many of the other parameters return values compatible with previous observations or theoretical work. This suggests the spectrum, even at this relatively low accretion rate, can be explained without the need for disc truncation. We find a preference for low spins over very high ($a>0.9$) spin. The spin could be higher if some modest amount of truncation occurs such that the disc terminates at $\approx 6-9$ $r_g$. We discuss some of the benefits and drawbacks of both our modelling approach and using \mx\ as a source. 

\section*{Acknowledgements}

This work has made use of observations obtained with \xmm, an ESA science mission with instruments and contributions directly funded by ESA Member States and NASA. This work also made use of observations from the \nustar\ mission, funded by NASA, and data supplied by UK \swift\ Science Data Centre and has made use of data and/or software provided by the High Energy Astrophysics Science Archive Research Center (HEASARC), which is a service of the Astrophysics Science Division at NASA/GSFC. This research utilized MAXI data provided by RIKEN, JAXA, and the MAXI team.
This research has made use of NASA’s Astrophysics Data System. SDD and ML acknowledge support from STFC studentships. This research used the ALICE High Performance Computing Facility at the University of Leicester. Finally, we would like to thank the anonymous referee for their detailed and constructive feedback, which has enhanced of the quality of our manuscript.

\section*{Data Availability}

The observational data underlying this paper are available from the \xmm\ Science
Archive (\url{https://nxsa.esac.esa.int/nxsa-web/#home}), the \nustar\ Archive (\url{https://heasarc.gsfc.nasa.gov/docs/nustar/nustar_archive.html}), the \swift\ Archive (\url{https://heasarc.gsfc.nasa.gov/cgi-bin/W3Browse/swift.pl}), the \swift/BAT Hard X-ray Transient Monitor (\url{https://swift.gsfc.nasa.gov/results/transients/}) and the  MAXI 
Source List (\url{http://maxi.riken.jp/top/slist.html}). Additional information will be shared on reasonable request to the authors.




\bibliographystyle{mnras}
\bibliography{MAXI_paper_references.bib}




\appendix

\section{Markov Chain Monte Carlo (MCMC)} 
\label{MCMC_Appendix}

Here we provide more details of our use of MCMC to sample from the posterior.  We used the Goodman-Weare method with the `stretch move' \citep{GW_2010}.

For the RGS fitting of Section~\ref{RGS_section}, we used $M=100$ walkers and ran the chain for $N=100$ iterations as the \emph{warm up} phase, followed by $N=500$ iterations as the \emph{sampling} phase. Following some numerical experiments, we found that running the chains for longer or using a longer warm up phase did not change the main posterior summaries (median, variance) significantly. The final output was $M\times N = 5\times 10^4$ samples from which we estimate the properties of the posterior. 

The EPIC pn and \nustar\ fitting was far more challenging. The model has a modestly high number of parameters ($>20$) and some parameters have narrow, informative priors while others have wide, non-informative priors. Some combinations of parameters are highly covariant in the fitting. We used $M=200$ walkers -- due to the larger number of parameters -- and ran the chains for a long warm up phase until our diagnostics indicated that further sampling would not change any key results. We then ran a $N=3 \times 10^4$ iteration sampling phase ($6\times 10^6$ samples in total) but \emph{thinned} the chains by a factor $600$ to yield a final set of $10^4$ posterior samples.

\subsection{Initialisation}

We first used interactive fitting to estimate the best fit (in terms of minimising $W$-stat for RGS data or $\chi^2$ for EPIC pn and \nustar\ data) and used this to initialise the MCMC. We initialised the walkers in a small `ball': each of the $M$ walkers' initial position was drawn from a narrow Gaussian centred on the best fit. Some parameters from our initial optimisation were near to boundaries of the parameter space (e.g. $A_{Fe}$ or $\log(\xi)$) and without further treatment, some of the random initial points would fall outside of the allowed ranges. We avoided this by drawing a large number of points and selected the first $M$ that lay within the allowed parameter space and using these as the starting point for the walkers.

\subsection{Warm up}

We computed the $\chi^2$ value for each sample parameters in order to track the evolution of the chains. We found that during warm up, the chain of walkers would sometimes divide into two or more separate clusters in $\chi^2$, presumably representing local minima in the  $\chi^2$-space. The lowest $\chi^2$ of these could be significantly lower than the others (meaning vastly larger posterior density). These regions could be widely separated in parameter space, relative to the size of the cluster of walkers exploring that region, meaning that walkers rarely jumped from one local attractor to a better one (higher posterior density). This is the problem of isolated local minima. 
We found that small clusters of walkers getting stuck with relatively high $\chi^2$ values was a common problem and a serious impediment to the warm up phase. 

We ran MCMC chains in batches of $N=10^4$ iterations per walker (with $M=200$ walkers this generated $2 \times 10^6$ samples per batch). After the first batch, we took measures to mitigate the issue of `trapped' walkers. We discarded $75$\% of points (for any walker/iteration along the chain) with highest $\chi^2$. We then randomly sampled $200$ points from the remaining points to use as initialisation for each walker in the next batch. We then ran more batches, using the end state of one batch to initialise the next -- effectively one long chain -- until none of our convergence diagnostics were changing significantly with additional batches. This signalled the end of the warm up phase.

One of our fits was treated differently. Examination of the first batch of samples from the spin free model fitted to obs3, showed only a single $\chi^2$ cluster, but the walkers had separated into three regions of parameter space with different volumes. 
In this case, we did not discard the $75$\% of samples with worst $\chi^2$ but instead used the end point of the first batch as the start of the second.

\subsection{Criteria for convergence} \label{Criteria_for_convergence}

Assessing convergence of MCMC chains in high dimensions, and from ensemble methods in particular, is challenging. 

Our first check is a visual inspection of the trace plots showing the path of the mean of the ensemble of $M$ walkers, and its variance, as a function of iteration number. We did this for every parameter after each batch of $N=10^4$ iterations, and we extended the warm up phase by another batch whenever any parameter showed visual signs of significant drift in the mean or variance. 

We quantified this by comparing changes in the mean of the ensemble to the variance within the ensemble. We took $1000$ iterations around iteration $i$ and another $1000$ around iteration $j$.
From these we computed the mean, $\mu$, and an estimate of the width, $S$ (we used difference between the $84$th and $16$th centiles).
We then used:
\begin{equation}
\sqrt{2} \frac{\mu_i - \mu_j}{(S_i^2 + S_j^2)^{1/2}}
\end{equation}
as a relative measure of the change in the mean from $i$ to $j$, relative to the width of the distribution. We did this for each parameter, comparing the movement between four locations in the chain: $9\times10^3$ - $1\times10^4$ iterations, $4.9\times10^4$ - $5.0\times10^4$ iterations, $9.9\times10^4$ - $1.0\times10^5$ iterations and $1.99\times10^5$ - $2.00\times10^5$ iterations. 
Between the final two locations the change in the mean was $< 10\%$ of the standard deviation for almost all parameters (on average, the parameter means changed by $4.06\%$ of the standard deviation), as expected for a chain near convergence.

As a check of whether the final batch of samples from the MCMC are drawn from the target posterior distribution, we compared them to the NS samples. MCMC and NS are different methods for sampling from a posterior distribution and a close agreement between the two is suggestive of the convergence a both. We found good agreement in the cases of obs1 and obs2, but not for obs3. The posterior samples generated from MCMC and NS were not consistent even after very long warm up phases for the MCMC.

\subsection{Autocorrelation}

Despite an extensive warm up phase to allow the chain to stabilise, our chains -- like all MCMC output -- suffers from autocorrelation: new points in the chain are correlated with earlier points in the chain. This can lead poor sampling of the posterior and biased parameter inferences if the sampling phase is shorter than the typical autocorrelation time. For the Goodman-Weare method, the conventional way to assess autocorrelation of the ensemble is with the integrated autocorrelation time ($\tau$), discussed in \cite{GW_2010}. 

We follow the advice of Foreman-Mackay et al.\footnote{\url{https://emcee.readthedocs.io/en/stable/tutorials/autocorr/}} and estimate the integrated autocorrelation time for each walker, and average this over all walkers to provide an initial measure of $\tau$. We also fitted a Gaussian Process model (specifically, second-order continuous autoregressive model, CAR(2)) simultaneously to each walker using the \texttt{celerite} software \citep{Foreman-Mackey_2017_celerite}. The model parameters can then be converted into an autocorrelation time, $\tau$. We do this for each of the parameters in our model.

The autocorrelation time can itself be difficult to estimate unless the chains are many times longer than $\tau$. We repeat the estimation of $\tau$ (both methods) for different lengths of chain, $N$, and look for signs that the estimates are converging as the chain length increases.  
By extrapolating the trend of $\tau$ with $N$, we estimate most parameters should have an autocorrelation length of $\sim$ $10^4$ iterations. A few parameters appeared to show a much longer autocorrelation time, $\gtrsim$ $10^5$. After discarding the samples from the warm up phase, some parameters may have only run for one autocorrelation length even with our long chains, meaning and there would be only one independent value per walker over the whole chain. But with $200$ walkers we have an effective sample size of $\sim 200$ even in this worst case. 

\section{Nested sampling (NS)} 
\label{NS_Appendix}

For the EPIC pn and \nustar\ spectra, we preformed NS on each of the observations and for the spin low, high and free models. We found convergence to be too slow using the MLfriends algorithm to sample new `live' points \citep{Buchner_2014, Buchner_2019}. In order to reduce the computational cost of convergence, we changed to the `step sampler' (also known as the slice sampler; \citealp{Buchner_2022}) after an initial run of each model. We tested differing numbers of steps ($20 - 200$) and found the evidence ($\ln(Z)$), $W$-stat and posterior distributions were stable for $\geq 80$ steps. We started the algorithm with $400$ live points.

\section{The Bayes Factor (BF)} 
\label{Analysis_Methods}

The Bayes Factor (BF) is a way to compare two models in terms of the ratio of their marginal likelihoods (sometimes called {\it evidences}). It represents the factor by which we update our prior odds on a pair of models as a result of analysing the data. A Bayes factor $BF \sim 1$ indicates no strong evidence to favour either model, whereas a very high or low BF indicates a strong preference for one model over the other. The BF is often reported on a logarithmic scale, with $\log(BF) \sim 0$ indicating no preference. 
\cite{Practical_Bayesian_Inference_Book} suggest that with a Bayes Factor in the range $0.1-10$ (odds of $10:1$ or less in favour of one model) there is insufficient evidence to prefer one model over the other, i.e. with $-1 \lesssim \log(BF) \lesssim 1$ (or in the range $-1.7 - 1.7$ to be conservative, corresponding to odds of less than $50:1$).

There are several ways to estimate Bayes Factors; here, we use two:
\begin{enumerate}
  \item Nested sampling returns a direct estimate of the marginal likelihood for each model in terms of $\ln(Z)$. The difference of the $\ln(Z)$ estimates for any pair of models gives $\ln(BF)$. 
  \item The Savage-Dickey Density Ratio (SDDR)  provides an simple way to compute the BF from posterior density estimates (such as from MCMC or NS) for nested models under certain conditions \citep{Verdinelli_1995, Trotta_2007, Wagenmakers_2010}.
  We consider two models: model $M_0$ has spin parameter fixed at some value $a=a^*$ and model $M_1$ has $a$ as a free parameter. $M_0$ is nested within the more general model $M_1$, and the priors on all other parameters are the same between the two models and independent of the value of the $a$ parameter. The SDDR is the ratio of the marginal posterior density to the marginal prior density, at $a=a^*$, computed from $M_1$ and this is exactly equivalent to the BF for $M_0$ against $M_1$, $BF_{01}$, under the above conditions. Given posterior samples from our free spin model (with either MCMC or NS), we can estimate the marginal posterior for $a$ and use this to compare posterior and prior density at any value $a=a^*$. We computed two SDDR values, one for $a^* = 0$ (zero-spin) and one for $a^* = 0.998$ (maximal spin). 
\end{enumerate}

\section{The typical set}
\label{sect:typical}

The idea of a \emph{typical set} for some probability density function comes from information theory. Loosely, if we were drawing random samples from a density function supported over some space (e.g. a posterior), the typical set is the region of the space where most of the draws come from. Put another way, the typical set is the region of the space that contains most of the probability mass. Importantly, and perhaps surprisingly, this often does not include the peak of the probability density. For high dimensional posterior distributions (e.g. more than a few parameters) the typical set can be a relative thin region around but at some distance from the posterior mode\footnote{See this discussion of typical sets for further explanation: \url{https://mc-stan.org/users/documentation/case-studies/curse-dims.html}.}.
In practice, this means that samples from the typical set will have a lower posterior density (e.g. higher Bayesian $\chi^2$) than a sample near or at the posterior mode/best fit. If an MCMC sampler is working well, then most of the samples drawn should be from the typical set and we should expect most samples to correspond to a $\chi^2$ that is somewhat worse than the best possible.
See \citet{Nalisnick_2019} for a more detailed test of `typicality' of random samples based on similar ideas.

We can get a very rough idea of the magnitude of this effect -- the difference between the best fit and a fit taken from the typical set -- by considering the following calculation.  

Consider $D$ random variables that are independently and identically distributed following a standard normal distribution. The joint probability density function is a ($D$-dimensional) multivariate Gaussian. If $r$ is the Euclidean distance from the origin, then the mode lies at $r=0$. What is the distance from the mode to a typical point? E.g. the average of $r$. The variable $r^2$ will have a chi-square distribution with $D$ degrees of freedom (dof), and $r$ will have a \emph{chi} distribution with $D$ dof. The average distance, $\langle r \rangle$ will equal the mean from the chi distribution, $\mu_D$. For $D=24$ this is $\mu_D \approx 4.8$. This means that `typically’ a random draw from a $D=24$ dimensional Gaussian distribution is $\approx 4.8\sigma$ from the mode. From the properties of the multivariate Gaussian, the difference in log density at the mode and at a point $4.8\sigma$ from the mode is $-4.8^2/2$. The $\chi^2$ statistic is related to the log density by $\ln(p) = -\chi^2/2$ (ignoring constant terms) and therefore the corresponding $\Delta \chi^2 \approx +23$. In words: when sampling a $D=24$ dimensional posterior, a set of parameter values picked randomly from the typical set could be expected to give a fit that is worse than the best fit (minimum $\chi^2$) by $\Delta \chi^2 \approx +23$. This may be a surprising result, but derives from the basic features of probability densities in high dimensional spaces.

In our Bayesian modelling problem presented above, the posterior is not a multivariate Gaussian, but it does appear unimodal and relatively simple in shape and so a Gaussian may be a reasonable approximation near the mode. We can therefore expect that if the MCMC is working well, and sampling from the posterior, the $\chi^2$ of samples should be worse than the best fit by $\sim 23$. We used this as another `rule of thumb' for judging the convergence of the MCMCs used above. By optimising we can estimate the $\chi^2$ at the best fit, and we can compute the $\chi^2$ statistic for the MCMC samples. If the latter are too close or too far from the former (compared to $\sim 23$), we assume the sampler is not yet drawing from the typical set and so required more iterations to warm up.

\section{MCMC and NS Tables}

\begingroup
\begin{landscape}
\renewcommand{\arraystretch}{1.4}
\begin{table}
	\centering
	\caption{MCMC results for \xmm\ EPIC pn and \nustar\ data. Values are given in terms of the median, with errors defining the range from the 16th to 84th quantiles. Units are as follows: $10^{22}$ cm$^{-2}$ for $N_H$; $10^{16}$ cm$^{-2}$ for other column densities; degrees for $i$; \Msun\ for $M_{BH}$; 10$^{18}$ g s$^{-1}$ for $\dot{M}$ and kpc for $D_{BH}$. }
	\label{tab:Param_values_MCMC}
	\begin{tabular}{l l c c c c c c c c c}
 	\hline
 	\hline
 	& & \multicolumn{3}{c}{obs1} & \multicolumn{3}{c}{obs2} & \multicolumn{3}{c}{obs3}\\
	Model & Parameter & a = 0 & a = 0.998 &  a = free & a = 0 & a = 0.998 &  a = free & a = 0 & a = 0.998 &  a = free\\

\hline
\hline
\textsc{ISMabs} & $N_H$ & 0.095$^{+0.015}_{-0.014}$ & 0.091$^{+0.014}_{-0.014}$ & 0.093$^{+0.015}_{-0.014}$ & 0.115$^{+0.014}_{-0.014}$ & 0.112$^{+0.014}_{-0.014}$ & 0.115$^{+0.015}_{-0.015}$ & 0.134$^{+0.013}_{-0.012}$ & 0.132$^{+0.013}_{-0.013}$ & 0.133$^{+0.012}_{-0.012}$\\
& $N_{OI}$ & 80.337$^{+2.369}_{-2.375}$ & 80.120$^{+2.342}_{-2.408}$ & 80.102$^{+2.343}_{-2.428}$ & 80.889$^{+2.269}_{-2.333}$ & 80.541$^{+2.329}_{-2.334}$ & 80.813$^{+2.249}_{-2.213}$ & 81.278$^{+2.353}_{-2.327}$ & 81.414$^{+2.321}_{-2.433}$ & 81.295$^{+2.268}_{-2.242}$\\
& $N_{OII}$ & 3.069$^{+0.747}_{-0.724}$ & 2.989$^{+0.742}_{-0.735}$ & 3.028$^{+0.708}_{-0.757}$ & 3.042$^{+0.743}_{-0.745}$ & 2.992$^{+0.750}_{-0.734}$ & 3.005$^{+0.773}_{-0.713}$ & 3.082$^{+0.754}_{-0.707}$ & 3.066$^{+0.754}_{-0.755}$ & 3.084$^{+0.699}_{-0.693}$\\
& $N_{OIII}$ & 0.038$^{+0.031}_{-0.025}$ & 0.038$^{+0.033}_{-0.025}$ & 0.039$^{+0.034}_{-0.025}$ & 0.038$^{+0.033}_{-0.025}$ & 0.038$^{+0.034}_{-0.025}$ & 0.036$^{+0.034}_{-0.024}$ & 0.036$^{+0.036}_{-0.024}$ & 0.038$^{+0.034}_{-0.025}$ & 0.039$^{+0.030}_{-0.025}$\\
& $N_{NeI}$ & 9.545$^{+0.971}_{-0.951}$ & 9.550$^{+0.970}_{-0.961}$ & 9.485$^{+0.969}_{-0.987}$ & 9.408$^{+0.987}_{-0.977}$ & 9.537$^{+0.961}_{-0.983}$ & 9.423$^{+0.991}_{-0.974}$ & 9.505$^{+0.986}_{-0.996}$ & 9.585$^{+0.994}_{-0.952}$ & 9.457$^{+1.033}_{-0.948}$\\
& $N_{NeII}$ & 0.004$^{+0.004}_{-0.003}$ & 0.004$^{+0.004}_{-0.003}$ & 0.004$^{+0.004}_{-0.003}$ & 0.004$^{+0.004}_{-0.003}$ & 0.004$^{+0.004}_{-0.003}$ & 0.004$^{+0.004}_{-0.003}$ & 0.004$^{+0.004}_{-0.003}$ & 0.004$^{+0.004}_{-0.003}$ & 0.004$^{+0.004}_{-0.002}$\\
& $N_{NeIII}$ & 0.053$^{+0.054}_{-0.037}$ & 0.054$^{+0.053}_{-0.037}$ & 0.056$^{+0.057}_{-0.038}$ & 0.056$^{+0.053}_{-0.039}$ & 0.054$^{+0.053}_{-0.037}$ & 0.055$^{+0.056}_{-0.038}$ & 0.054$^{+0.053}_{-0.037}$ & 0.054$^{+0.055}_{-0.037}$ & 0.056$^{+0.055}_{-0.038}$\\
& $N_{Fe}$ & 3.570$^{+0.609}_{-0.603}$ & 3.588$^{+0.561}_{-0.579}$ & 3.613$^{+0.548}_{-0.602}$ & 3.875$^{+0.553}_{-0.576}$ & 3.878$^{+0.557}_{-0.567}$ & 3.821$^{+0.547}_{-0.565}$ & 4.103$^{+0.567}_{-0.547}$ & 4.080$^{+0.560}_{-0.561}$ & 4.090$^{+0.580}_{-0.523}$\\
\hline
\textsc{SIMPL} & $\Gamma_{S}$ & 1.670$^{+0.010}_{-0.010}$ & 1.666$^{+0.011}_{-0.010}$ & 1.669$^{+0.011}_{-0.009}$ & 1.692$^{+0.004}_{-0.004}$ & 1.691$^{+0.004}_{-0.004}$ & 1.691$^{+0.004}_{-0.004}$ & 1.863$^{+0.005}_{-0.005}$ & 1.862$^{+0.005}_{-0.005}$ & 1.862$^{+0.005}_{-0.005}$\\
& $f_{scat}$ & 0.709$^{+0.037}_{-0.036}$ & 0.728$^{+0.035}_{-0.035}$ & 0.713$^{+0.037}_{-0.037}$ & 0.730$^{+0.056}_{-0.052}$ & 0.763$^{+0.047}_{-0.048}$ & 0.732$^{+0.054}_{-0.060}$ & 0.876$^{+0.090}_{-0.156}$ & 0.932$^{+0.046}_{-0.065}$ & 0.863$^{+0.096}_{-0.172}$\\
\hline
\textsc{KERRBB} & $a$ & - & - & -0.162$^{+0.595}_{-0.554}$ & - & - & -0.396$^{+0.661}_{-0.432}$ & - & - & -0.419$^{+0.693}_{-0.419}$\\
& $i$ & 63.139$^{+2.991}_{-3.111}$ & 61.025$^{+2.830}_{-2.916}$ & 63.286$^{+2.824}_{-3.356}$ & 61.909$^{+2.909}_{-3.194}$ & 58.790$^{+3.124}_{-2.996}$ & 61.853$^{+3.068}_{-3.194}$ & 62.583$^{+3.292}_{-2.459}$ & 61.541$^{+2.673}_{-2.511}$ & 62.538$^{+3.212}_{-2.455}$\\
& $M_{BH}$ & 8.444$^{+0.786}_{-0.706}$ & 8.654$^{+0.820}_{-0.684}$ & 8.457$^{+0.828}_{-0.706}$ & 8.425$^{+0.835}_{-0.700}$ & 8.705$^{+0.826}_{-0.746}$ & 8.410$^{+0.771}_{-0.642}$ & 8.481$^{+0.759}_{-0.683}$ & 8.555$^{+0.846}_{-0.717}$ & 8.429$^{+0.825}_{-0.718}$\\
& $\dot{M} (\times 10^{-3})$ & 5.212$^{+2.324}_{-1.674}$ & 0.590$^{+0.217}_{-0.184}$ & 5.311$^{+3.316}_{-2.189}$ & 3.891$^{+1.678}_{-1.110}$ & 0.446$^{+0.152}_{-0.125}$ & 4.594$^{+2.638}_{-1.845}$ & 0.571$^{+0.254}_{-0.175}$ & 0.082$^{+0.034}_{-0.025}$ & 0.677$^{+0.411}_{-0.30}$\\
& $D_{BH}$ & 2.978$^{+0.558}_{-0.476}$ & 2.598$^{+0.444}_{-0.451}$ & 2.952$^{+0.538}_{-0.486}$ & 2.940$^{+0.509}_{-0.466}$ & 2.522$^{+0.417}_{-0.382}$ & 2.981$^{+0.531}_{-0.466}$ & 2.891$^{+0.542}_{-0.461}$ & 2.858$^{+0.530}_{-0.447}$ & 2.896$^{+0.578}_{-0.507}$\\
& $f_{col}$ & 2.207$^{+0.281}_{-0.282}$ & 1.125$^{+0.148}_{-0.089}$ & 2.273$^{+0.484}_{-0.467}$ & 2.034$^{+0.251}_{-0.246}$ & 1.093$^{+0.120}_{-0.066}$ & 2.171$^{+0.408}_{-0.428}$ & 1.900$^{+0.843}_{-0.635}$ & 1.389$^{+0.271}_{-0.227}$ & 1.863$^{+0.848}_{-0.603}$\\
\hline
\textsc{relxillCp} & $I_1$ & 2.841$^{+0.740}_{-0.656}$ & 2.412$^{+0.527}_{-0.537}$ & 2.887$^{+0.785}_{-0.704}$ & 2.632$^{+0.688}_{-0.615}$ & 2.511$^{+0.649}_{-0.592}$ & 2.749$^{+0.748}_{-0.632}$ & 2.704$^{+0.668}_{-0.633}$ & 2.432$^{+0.635}_{-0.543}$ & 2.793$^{+0.828}_{-0.692}$\\
& $I_2$ & 1.734$^{+0.204}_{-0.235}$ & 1.664$^{+0.220}_{-0.238}$ & 1.752$^{+0.187}_{-0.242}$ & 1.465$^{+0.155}_{-0.174}$ & 1.413$^{+0.175}_{-0.190}$ & 1.476$^{+0.153}_{-0.168}$ & 1.263$^{+0.215}_{-0.216}$ & 1.218$^{+0.228}_{-0.224}$ & 1.283$^{+0.221}_{-0.213}$\\
& $\Gamma_{R}$ & 1.371$^{+0.185}_{-0.112}$ & 1.421$^{+0.203}_{-0.130}$ & 1.368$^{+0.182}_{-0.106}$ & 1.911$^{+0.039}_{-0.040}$ & 1.937$^{+0.036}_{-0.036}$ & 1.914$^{+0.038}_{-0.043}$ & 2.209$^{+0.040}_{-0.027}$ & 2.218$^{+0.045}_{-0.028}$ & 2.209$^{+0.039}_{-0.029}$\\
& $\log(\xi)$ & 1.421$^{+0.402}_{-0.926}$ & 1.499$^{+0.353}_{-0.635}$ & 1.498$^{+0.353}_{-0.792}$ & 0.238$^{+0.182}_{-0.155}$ & 0.245$^{+0.182}_{-0.157}$ & 0.260$^{+0.171}_{-0.164}$ & 0.051$^{+0.090}_{-0.039}$ & 0.056$^{+0.109}_{-0.043}$ & 0.053$^{+0.114}_{-0.041}$\\
& $A_{Fe}$ & 7.904$^{+1.503}_{-2.197}$ & 7.884$^{+1.576}_{-2.257}$ & 7.894$^{+1.471}_{-2.225}$ & 9.419$^{+0.414}_{-0.711}$ & 9.569$^{+0.316}_{-0.565}$ & 9.475$^{+0.382}_{-0.660}$ & 9.460$^{+0.398}_{-0.804}$ & 9.471$^{+0.398}_{-0.832}$ & 9.510$^{+0.368}_{-0.896}$\\
& $norm_R (\times 10^{-3})$ & 1.428$^{+1.468}_{-0.683}$ & 1.049$^{+1.101}_{-0.498}$ & 1.425$^{+1.457}_{-0.697}$ & 0.522$^{+0.064}_{-0.053}$ & 0.515$^{+0.060}_{-0.052}$ & 0.518$^{+0.064}_{-0.055}$ & 0.069$^{+0.011}_{-0.009}$ & 0.072$^{+0.011}_{-0.011}$ & 0.069$^{+0.010}_{-0.010}$\\
\hline
\textsc{constant} & $C$ - FPM-A & - & - & - & 1.631$^{+0.005}_{-0.005}$ & 1.630$^{+0.005}_{-0.005}$ & 1.630$^{+0.005}_{-0.005}$ & 1.151$^{+0.007}_{-0.006}$ & 1.151$^{+0.006}_{-0.006}$ & 1.151$^{+0.006}_{-0.007}$\\
& $C$ - FPM-B & - & - & - & 1.617$^{+0.005}_{-0.004}$ & 1.616$^{+0.004}_{-0.004}$ & 1.617$^{+0.005}_{-0.005}$ & 1.159$^{+0.006}_{-0.006}$ & 1.158$^{+0.006}_{-0.007}$ & 1.159$^{+0.006}_{-0.007}$\\
\hline
\hline
    
	\end{tabular}
\end{table}
\end{landscape}
\endgroup

\begingroup
\begin{landscape}
\renewcommand{\arraystretch}{1.4}
\begin{table}
	\centering
	\caption{Nested sampling results for \xmm\ EPIC pn and \nustar\ data. Values are given in terms of the median, with errors defining the range from the 16th to 84th quantiles. Units are given in Table \ref{tab:Param_values_MCMC}.}
	\label{tab:Param_values_NS}
	\begin{tabular}{l l c c c c c c c c c}
 	\hline
 	\hline
 	& & \multicolumn{3}{c}{obs1} & \multicolumn{3}{c}{obs2} & \multicolumn{3}{c}{obs3}\\
	Model & Parameter & a = 0 & a = 0.998 &  a = free & a = 0 & a = 0.998 &  a = free & a = 0 & a = 0.998 &  a = free\\

\hline
\hline
\textsc{ISMabs} & $N_H$ & 0.097$^{+0.013}_{-0.013}$ & 0.092$^{+0.013}_{-0.013}$ & 0.093$^{+0.013}_{-0.013}$ & 0.116$^{+0.013}_{-0.015}$ & 0.113$^{+0.015}_{-0.016}$ & 0.112$^{+0.011}_{-0.011}$ & 0.085$^{+0.012}_{-0.011}$ & 0.095$^{+0.012}_{-0.008}$ & 0.095$^{+0.014}_{-0.013}$\\
& $N_{OI}$ & 81.139$^{+2.046}_{-1.912}$ & 79.918$^{+2.151}_{-2.082}$ & 80.666$^{+1.991}_{-1.755}$ & 80.506$^{+2.366}_{-2.325}$ & 80.997$^{+1.851}_{-2.20}$ & 80.426$^{+2.130}_{-1.684}$ & 80.055$^{+1.953}_{-1.976}$ & 80.427$^{+1.557}_{-1.471}$ & 80.467$^{+1.893}_{-1.946}$\\
& $N_{OII}$ & 3.097$^{+0.692}_{-0.639}$ & 2.629$^{+0.516}_{-0.583}$ & 3.062$^{+0.621}_{-0.70}$ & 3.029$^{+0.668}_{-0.677}$ & 3.204$^{+0.644}_{-0.754}$ & 3.263$^{+0.617}_{-0.663}$ & 3.022$^{+0.638}_{-0.619}$ & 3.366$^{+0.556}_{-0.617}$ & 2.950$^{+0.745}_{-0.691}$\\
& $N_{OIII}$ & 0.038$^{+0.036}_{-0.026}$ & 0.035$^{+0.031}_{-0.021}$ & 0.039$^{+0.031}_{-0.023}$ & 0.038$^{+0.032}_{-0.024}$ & 0.033$^{+0.024}_{-0.018}$ & 0.036$^{+0.029}_{-0.021}$ & 0.026$^{+0.024}_{-0.017}$ & 0.033$^{+0.021}_{-0.018}$ & 0.048$^{+0.029}_{-0.023}$\\
& $N_{NeI}$ & 9.663$^{+0.942}_{-0.906}$ & 9.737$^{+0.872}_{-0.894}$ & 9.424$^{+0.833}_{-0.833}$ & 9.357$^{+0.988}_{-0.972}$ & 9.369$^{+1.083}_{-0.946}$ & 9.918$^{+0.722}_{-0.761}$ & 9.313$^{+0.801}_{-0.864}$ & 9.564$^{+0.919}_{-0.662}$ & 9.119$^{+0.702}_{-0.783}$\\
& $N_{NeII}$ & 0.005$^{+0.003}_{-0.002}$ & 0.004$^{+0.004}_{-0.003}$ & 0.003$^{+0.003}_{-0.002}$ & 0.005$^{+0.003}_{-0.003}$ & 0.004$^{+0.003}_{-0.003}$ & 0.004$^{+0.003}_{-0.002}$ & 0.004$^{+0.003}_{-0.003}$ & 0.004$^{+0.004}_{-0.003}$ & 0.004$^{+0.004}_{-0.002}$\\
& $N_{NeIII}$ & 0.052$^{+0.054}_{-0.034}$ & 0.055$^{+0.051}_{-0.035}$ & 0.056$^{+0.045}_{-0.035}$ & 0.046$^{+0.047}_{-0.031}$ & 0.036$^{+0.050}_{-0.026}$ & 0.042$^{+0.045}_{-0.028}$ & 0.024$^{+0.030}_{-0.016}$ & 0.040$^{+0.029}_{-0.022}$ & 0.057$^{+0.048}_{-0.038}$\\
& $N_{Fe}$ & 3.555$^{+0.560}_{-0.564}$ & 3.407$^{+0.579}_{-0.518}$ & 3.504$^{+0.542}_{-0.518}$ & 3.825$^{+0.558}_{-0.502}$ & 3.557$^{+0.637}_{-0.418}$ & 3.872$^{+0.523}_{-0.516}$ & 3.756$^{+0.459}_{-0.407}$ & 3.407$^{+0.463}_{-0.456}$ & 3.689$^{+0.538}_{-0.563}$\\
\hline
\textsc{SIMPL} & $\Gamma_{S}$ & 1.668$^{+0.011}_{-0.009}$ & 1.664$^{+0.010}_{-0.009}$ & 1.668$^{+0.009}_{-0.009}$ & 1.691$^{+0.003}_{-0.003}$ & 1.690$^{+0.004}_{-0.004}$ & 1.691$^{+0.003}_{-0.003}$ & 2.835$^{+0.324}_{-0.325}$ & 1.825$^{+0.054}_{-0.038}$ & 1.837$^{+0.021}_{-0.019}$\\
& $f_{scat}$ & 0.621$^{+0.030}_{-0.024}$ & 0.643$^{+0.028}_{-0.028}$ & 0.634$^{+0.031}_{-0.031}$ & 0.640$^{+0.056}_{-0.045}$ & 0.678$^{+0.050}_{-0.047}$ & 0.644$^{+0.043}_{-0.039}$ & 0.545$^{+0.20}_{-0.132}$ & 0.565$^{+0.042}_{-0.037}$ & 0.582$^{+0.042}_{-0.041}$\\
\hline
\textsc{KERRBB} & $a$ & - & - & -0.407$^{+0.549}_{-0.396}$ & - & - & -0.141$^{+0.492}_{-0.411}$ & - & - & -0.251$^{+0.350}_{-0.373}$\\
& $i$ & 62.396$^{+2.293}_{-2.226}$ & 59.663$^{+1.784}_{-2.129}$ & 63.349$^{+2.583}_{-2.846}$ & 62.077$^{+2.207}_{-2.685}$ & 59.012$^{+2.412}_{-2.452}$ & 62.432$^{+2.449}_{-2.132}$ & 61.438$^{+1.867}_{-2.241}$ & 61.729$^{+1.993}_{-2.650}$ & 62.440$^{+2.254}_{-2.358}$\\
& $M_{BH}$ & 8.020$^{+0.633}_{-0.472}$ & 8.624$^{+0.751}_{-0.601}$ & 8.091$^{+0.644}_{-0.512}$ & 8.511$^{+0.737}_{-0.717}$ & 8.879$^{+0.826}_{-0.644}$ & 8.396$^{+0.574}_{-0.612}$ & 8.368$^{+0.663}_{-0.609}$ & 8.302$^{+0.770}_{-0.524}$ & 8.310$^{+0.659}_{-0.645}$\\
& $\dot{M} (\times 10^{-3})$ & 4.903$^{+1.378}_{-1.659}$ & 0.674$^{+0.142}_{-0.098}$ & 6.902$^{+2.452}_{-2.059}$ & 3.813$^{+0.999}_{-0.658}$ & 0.471$^{+0.125}_{-0.049}$ & 4.058$^{+1.205}_{-1.196}$ & 0.341$^{+0.125}_{-0.10}$ & 0.065$^{+0.016}_{-0.026}$ & 0.504$^{+0.158}_{-0.124}$\\
& $D_{BH}$ & 3.002$^{+0.367}_{-0.499}$ & 2.860$^{+0.282}_{-0.211}$ & 3.210$^{+0.385}_{-0.287}$ & 2.989$^{+0.347}_{-0.319}$ & 2.696$^{+0.321}_{-0.194}$ & 2.941$^{+0.382}_{-0.316}$ & 3.233$^{+0.408}_{-0.474}$ & 2.675$^{+0.302}_{-0.599}$ & 2.666$^{+0.183}_{-0.191}$\\
& $f_{col}$ & 2.125$^{+0.312}_{-0.188}$ & 1.096$^{+0.097}_{-0.066}$ & 2.264$^{+0.317}_{-0.294}$ & 1.984$^{+0.210}_{-0.189}$ & 1.057$^{+0.073}_{-0.041}$ & 2.010$^{+0.311}_{-0.329}$ & 3.397$^{+0.689}_{-0.577}$ & 1.771$^{+0.223}_{-0.198}$ & 3.695$^{+0.489}_{-0.488}$\\
\hline
\textsc{relxillCp} & $I_1$ & 2.745$^{+0.718}_{-0.658}$ & 2.442$^{+0.515}_{-0.533}$ & 2.879$^{+0.695}_{-0.593}$ & 2.584$^{+0.550}_{-0.557}$ & 2.424$^{+0.436}_{-0.513}$ & 2.805$^{+0.696}_{-0.684}$ & 2.387$^{+0.492}_{-0.512}$ & 3.958$^{+0.892}_{-0.888}$ & 2.861$^{+0.763}_{-0.673}$\\
& $I_2$ & 1.747$^{+0.181}_{-0.199}$ & 1.692$^{+0.204}_{-0.223}$ & 1.771$^{+0.20}_{-0.224}$ & 1.458$^{+0.155}_{-0.168}$ & 1.426$^{+0.143}_{-0.147}$ & 1.449$^{+0.128}_{-0.176}$ & 3.578$^{+0.776}_{-0.807}$ & 3.407$^{+0.832}_{-0.792}$ & 2.968$^{+0.885}_{-0.804}$\\
& $\Gamma_{R}$ & 1.364$^{+0.167}_{-0.094}$ & 1.388$^{+0.180}_{-0.108}$ & 1.376$^{+0.155}_{-0.104}$ & 1.894$^{+0.037}_{-0.037}$ & 1.920$^{+0.035}_{-0.034}$ & 1.886$^{+0.035}_{-0.032}$ & 1.690$^{+0.012}_{-0.014}$ & 1.760$^{+0.106}_{-0.071}$ & 1.781$^{+0.072}_{-0.063}$\\
& $\log(\xi)$ & 1.539$^{+0.286}_{-0.563}$ & 1.622$^{+0.280}_{-0.432}$ & 1.521$^{+0.321}_{-0.539}$ & 0.256$^{+0.190}_{-0.159}$ & 0.265$^{+0.213}_{-0.163}$ & 0.273$^{+0.157}_{-0.159}$ & 4.054$^{+0.051}_{-0.041}$ & 4.327$^{+0.133}_{-0.141}$ & 4.345$^{+0.144}_{-0.115}$\\
& $A_{Fe}$ & 8.049$^{+1.450}_{-2.143}$ & 7.851$^{+1.537}_{-2.302}$ & 7.218$^{+2.047}_{-1.999}$ & 9.550$^{+0.333}_{-0.535}$ & 9.647$^{+0.254}_{-0.529}$ & 9.485$^{+0.371}_{-0.564}$ & 1.288$^{+0.447}_{-0.304}$ & 5.904$^{+2.235}_{-2.304}$ & 6.792$^{+2.117}_{-2.569}$\\
& $norm_R (\times 10^{-3})$ & 1.404$^{+1.308}_{-0.741}$ & 1.103$^{+0.974}_{-0.518}$ & 1.359$^{+1.302}_{-0.619}$ & 0.514$^{+0.053}_{-0.043}$ & 0.503$^{+0.051}_{-0.043}$ & 0.517$^{+0.046}_{-0.044}$ & 0.222$^{+0.015}_{-0.017}$ & 0.067$^{+0.026}_{-0.019}$ & 0.044$^{+0.012}_{-0.011}$\\
\hline
\textsc{constant} & $C$ - FPM-A & - & - & - & 1.630$^{+0.005}_{-0.005}$ & 1.629$^{+0.005}_{-0.005}$ & 1.630$^{+0.004}_{-0.004}$ & 1.156$^{+0.006}_{-0.006}$ & 1.156$^{+0.006}_{-0.006}$ & 1.154$^{+0.006}_{-0.006}$\\
& $C$ - FPM-B & - & - & - & 1.616$^{+0.004}_{-0.004}$ & 1.616$^{+0.004}_{-0.004}$ & 1.616$^{+0.004}_{-0.004}$ & 1.163$^{+0.006}_{-0.006}$ & 1.162$^{+0.006}_{-0.006}$ & 1.162$^{+0.006}_{-0.006}$\\
\hline
\hline
    
	\end{tabular}
\end{table}
\end{landscape}
\endgroup


\bsp	
\label{lastpage}
\end{document}